\documentclass[12pt,aps,prb,preprint]{revtex4}   
\usepackage{amsmath}    
\usepackage{graphicx}   
\begin{document}
\title{ The coordinate-free approach to spherical harmonics}
\author{Miguel P\'{e}rez-Saborid}
 \affiliation{Escuela Superior de Ingenieros, Area de Mec\'{a}nica de Fluidos,
 Universidad de Sevilla}
 \email{psaborid@us.es}   
\date{\today}
\begin{abstract}
We present in a unified and self-contained manner the
coordinate-free approach to spherical harmonics initiated in the
mid 19th century by James Clerk Maxwell, William Thomson and Peter
Guthrie Tait. We stress the pedagogical advantages of this
approach which leads in a natural way to many physically relevant
results that students find often difficult to work out using
spherical coordinates and associated Legendre functions. It is
shown how most physically relevant results of the theory of
spherical harmonics - such as recursion relations, Legendre's
addition theorem,surface harmonics expansions, the method of
images, multipolar charge distributions, partial wave expansions,
Hobson's integral theorem, rotation matrix and Gaunt's integrals -
can be efficiently derived in a coordinate free fashion from a few
basic elements of the theory of solid and surface harmonics
discussed in the paper.
\end{abstract}
\maketitle
\section{Introduction}
Spherical harmonics present themselves in many branches of physics
and applied mathematics such as electromagnetism, quantum
mechanics and gravitation theory. In most modern textbooks dealing
with topics of mathematical physics \cite{But,Jack}, the theory of
spherical harmonics is built upon their representation in
spherical coordinates. As is well known, in this approach one
starts by expressing Laplace's equation in spherical polar
coordinates ($r$,$\theta$,$\varphi$), where $r$ is the distance to
the origin and $\theta$ and $\varphi$ are the polar and azimuthal
angles, respectively. One then tries a solution in separated
variables that yields the spherical harmonics in the form
$Y_{lm}(\theta,\varphi)= C_{lm} e^{im\varphi} P_l^m(\cos \theta)$
($l=0,1,..., -l\le m\le l$), where $C_{lm}$'s are normalization
constants and the $P_l^m(\mu)$'s are the associated Legendre
functions defined by $2^l l!
P_l^m(\mu)\equiv(-1)^m(1-\mu^2)^{m/2}d^{l+m}(\mu^2-1)^l/d\mu^{l+m}$
. However, this representation often leads to cumbersome
derivations of the main results needed in physical applications,
so that, from a physicist's point of view, it may be desirable to
have a more compact approach in which mathematical manipulations
become more direct and, perhaps, also more guided by physical
insight. This is also the case with other special functions of
great physical relevance such as Bessel's functions or ellipsoidal
harmonics.

 The first representations of spherical harmonics freed from
spherical coordinates were those given by Thomson and
Tait\cite{TT} in their {\it Treatise on Natural Philosophy} and by
Maxwell\cite{Max} in his {\it Treatise on Electricity and
Magnetism}. The principle of their method is that, having given
any solution of Laplace's equation, other solutions may be
obtained by differentiating the given solution any number of times
in the directions of the cartesian axes $x$,$y$ and $z$ or, as
Maxwell proposed, in any directions whatsoever. In particular, by
applying this method to the simple solution $1/r$, with
$r=\sqrt{x^2+y^2+z^2}$, all the system of spherical harmonics may
be generated\cite{Hob}. Maxwell\cite{Max} also contributed with an
important integral theorem which apparently has not deserved by
itself much attention in the literature. In fact, after it first
appeared in Maxwell's treatise, it has almost invariably been
presented as a corollary of a more general - but also more
complicated - integral theorem first introduced by
Niven\cite{Niven} and generalized afterwards by Hobson\cite{Hob}.
Nevertheless, we consider Maxwell's theorem as an essential
ingredient in our approach and, therefore, have preferred to
follow here a different route that leads to Hobson's theorem via
Maxwell's theorem.

The aim of this paper is to show how the contributions due to
Maxwell and to Thomson and Tait lead to a unified, self-consistent
and spherical coordinate-free approach through which most of the
properties of spherical harmonics needed in usual physical
applications can easily and naturally be derived. We hope that
this paper helps students and researchers to understand and work
out by themselves many of the results they need when dealing with
specific problems involving spherical harmonics, thus avoiding
much of the trouble of constantly having to look them up in the
appendices of the relevant literature or in specialized
mathematical handbooks\cite{Abra}. For these purposes, we have
based our presentation on the simple properties of three basic
elements of the theory of solid and surface harmonics: a) the
elementary solid harmonics of the form $({\bf b\cdot r})^l$-$l$
being a nonnegative integer and ${\bf b}$ a null vector-, b)
Maxwell's harmonics, and c) Maxwell's integral theorem. We have
introduced these basic concepts in Section II and have applied
them in the rest of the paper to provide systematic,
coordinate-free derivations of some physically relevant results
such as the construction of the standard set of spherical
harmonics $Y_{lm}$ (Section III), recursion relations (Section
IV), Legendre's addition theorem (Section V), surface harmonics
expansions (Section VI) - with applications to the method of
images and multipolar charge distributions-, partial wave
expansions and Hobson's integral theorem (Section VII), rotation
matrix (Section VIII) and Gaunt's integrals (Section IX). Finally
conclusions are presented in Section X.

\section{General properties of solid and surface harmonics.}

We define a solid harmonic of the type $H_l$ or, more briefly, a
{\it $H_l$-harmonic}, as any solution of Laplace's equation,
\begin{equation}
\nabla^2 \phi=0, \label{ecu5}
\end{equation}
which is a homogenous polynomial of degree $l$ in the (cartesian)
variables ($x$, $y$, $z$):
\begin{equation}
H_l({\bf r})=\sum_{\alpha+\beta+\gamma=l} C_{\alpha \beta \gamma}
x^{\alpha} y^{\beta} z^{\gamma} ,\label{ecu6}
\end{equation}
where ${\bf r}=x {\bf e}_x + y {\bf e}_y + z {\bf e}_z$ is the
position vector and constants $C_{\alpha\beta\gamma}$ are
generally complex. As can be checked by direct substitution,  to
every $H_l$-harmonic there corresponds a second solution of
(\ref{ecu5}) of the form
\begin{equation}
V_l({\bf r})=\frac{H_l({\bf r})}{r^{2l+1}}, \label{ecu7}
\end{equation}
which we will define as a solid harmonic of the type $V_l$ or,
more briefly, {\it $V_l$-harmonic}. Note the following behaviors
of $H_l$ and $V_l$ for small and large values of $r$
\begin{equation}
r\rightarrow 0 :\quad  H_l({\bf r})\rightarrow 0,\quad V_l({\bf
r})\rightarrow \infty, \label{eq7a} \end{equation}
\begin{equation}
r\rightarrow \infty :\quad  H_l({\bf r})\rightarrow \infty,\quad
V_l({\bf r})\rightarrow 0. \label{eq7b} \end{equation} According
to (\ref{eq7a}), $H_l$-harmonics and $V_l$-harmonics are also
known as {\it regular} and {\it irregular} solid
harmonics\cite{Wenig2}, respectively.

A {\it surface harmonic} of degree $l$, $Y_l$, is defined as the
function obtained from a $H_l$-harmonic or, equivalently, a
$V_l$-harmonic according to the rules
\begin{equation}
Y_l({\bf r})=H_l({\bf r})/r^{l}=r^{l+1} V_l({\bf r}) .\label{ecu8}
\end{equation}
Since $H_l$ is a homogeneous polynomial of degree $l$ in the
variables $x$, $y$ and $z$, it is clear that the value of $Y_l$ at
any point of space, ${\bf r}$, is the same as that at the point of
the unit sphere ${\bf e}_r={\bf r}/r$: $Y_l({\bf r})=Y_l({\bf
e}_r)$. Also, note that the values of $Y_l$, $H_l$ and $V_l$
coincide on the unit sphere.

It can easily be shown \cite{TT, Sch} that the number of
independent $H_l$ - harmonics is $2l+1$ which, of course, must
also be the number of independent functions $V_l$ and $Y_l$ by
(\ref{ecu7}) and (\ref{ecu8}). In effect, to a given value of the
exponent $\alpha$ ($0\le \alpha \le l$) in a homogeneous
polynomial of degree $l$ such as (\ref{ecu6}) there correspond
$l-\alpha +1$ possible values of the exponent $\beta$, since the
value of $\gamma$ is fixed by $\gamma=l-\alpha-\beta$. Thus, the
maximum number of monomials of the form $x^\alpha y^\beta
z^\gamma$ in (\ref{ecu6}) will be $\sum_{\alpha=0}^{l}
(l-\alpha+1)=(l+1)(l+2)/2$. Since $\nabla^2 H_l$  is a homogeneous
polynomial of degree $l-2$, it will therefore contain a maximum of
$(l-1)l/2$ terms whence the condition $\nabla^2 H_l=0$ is
equivalent $l(l-1)/2$ equations between the constant coefficients
$C_{\alpha \beta \gamma}$  in (\ref{ecu6}). The number of
independent constant remaining is thus $(l+1)(l+2)/2-l(l-1)/2$, or
$2l+1$. Therefore, any solid or surface harmonic of degree $l$ can
be expressed as a linear combination of $2l+1$ independent
harmonics of the same class.

Observe that, unlike functions $H_l$ or $V_l$, surface harmonics
do not satisfy Laplace's equation, but instead
\begin{equation}
\nabla^2 Y_l + \frac{l(l+1)}{r^2} Y_l=0 , \label{ecu9}
\end{equation}
as is readily verified by direct substitution of (\ref{ecu8}) into
(\ref{ecu5}). From (\ref{ecu9}) it is easy to obtain the
orthogonality property\cite{TT} for two given surface harmonics,
$Y_{l}$ and $X_{n}$. In effect, if we substract the equation for
$Y_l$ multiplied by $X_n$ from that for $X_n$ multiplied by $Y_l$
and integrate over the volume of a sphere of radius $a$ we obtain
\begin{equation}
  \left[l(l+1)-n(n+1)\right] \int_{r < a} dV
\frac{1}{r^2} Y_l X_n= - \int_{r=a} dS \left(X_n {\bf e}_r\cdot
\nabla Y_l-Y_l {\bf e}_r\cdot \nabla X_n\right)=0, \label{ecu9aa}
\end{equation}
where the surface integral has been obtained through Gauss's
theorem on using the identity $X_n \nabla^2 Y_l-Y_l  \nabla^2
X_n=\nabla\cdot\left(X_n \nabla Y_l-Y_l \nabla X_n\right)$. Since
surface harmonics do not depend on $r$, the surface integral
vanishes and the volume integral in the LHS of (\ref{ecu9aa}) can
be transformed into one on the unit sphere by writing $dV=r^2 dr
d\Omega$, $d\Omega$ being the element of solid angle. Carrying out
the trivial integration over the radial coordinate (\ref{ecu9aa})
yields
\begin{equation}
\int d\Omega Y_l X_n= 0 \quad {\rm if}\quad l \ne n. \label{ecu9b}
\end{equation}

An interesting relation between the solid and surface harmonics
satisfying (\ref{ecu8}) is obtained if one considers the function
\begin{equation}
\Psi\equiv \frac{1}{2l+1}\left[a^{-l+1} H_l({\bf r})-{ a^{l+2}}
V_l({\bf r})\right],\label{ecu13A1}
\end{equation}
where $a$ is any positive constant. On using (\ref{ecu8}) to
express $V_l$ and $H_l$ in terms $Y_l$ and taking into account
that ${\bf e}_r \cdot \nabla Y_l=0$ it is seen that $\Psi$
vanishes at $r=a$ and satisfies the identity
\begin{equation}
Y_l\equiv \frac{\partial \Psi}{\partial r}\Big|_{r=a}=\left[{\bf
e}_r \cdot \nabla \Psi \right]_{r=a}. \label{ecu13A2}
\end{equation}
The physical meaning\cite{Fe} of this identity is that $a^{-l+1}
H_l({\bf r})/(2l+1)$ and $a^{l+2} V_l({\bf r})/(2l+1)$ yield the
potential for $r \le a$ and $r \ge a$, respectively, due to an
arbitrarily thin sphere of radius $a$ with a surface charge
density distribution given by $Y_l({\bf e}_r)/(4\pi)$ -  the jump
at the surface in the normal derivatives of the potential being
represented by (\ref{ecu13A2}) -. This expression is useful when
one considers the integral over the sphere $r=a$ of the product of
$Y_l$ times any regular function $\Phi$ that satisfies $\nabla^2
\Phi = 0$ inside the sphere. Using the identity
$\nabla\cdot(\Phi\nabla\Psi)=\nabla\cdot(\Psi\nabla\Phi)+ \Phi
\nabla^2 \Psi-\Psi \nabla^2 \Phi$ with  $\nabla^2 \Phi=0$, Gauss's
theorem yields
\begin{equation}
 \int_{r=a} dS \Phi Y_l = \int_{r<a} dV \Phi \nabla^2 \Psi=- \frac{a^{l+2}}{2l+1}\int_{r<a} dV \Phi \nabla^2
V_l, \label{ecu13A4}
\end{equation}
where we have taken into account in (\ref{ecu13A1}) that $H_l$ is
regular for $r < a$ and $\nabla^2 H_l=0$. In (\ref{ecu13A4})
$\nabla^2 V_l$ must be considered as a generalized function due to
its singularity at the origin.

\subsection{The simplest solid harmonics}

 The simplest $H_l$-harmonics are polynomials
 of the form
 $({\bf b}\cdot{\bf r})^l$, where ${\bf b}$ is any constant
(complex) null vector, i.e., ${\bf b}\cdot{\bf b}=0$,  so that
\begin{equation}
\nabla^2({\bf b}\cdot{\bf r})^l=l\nabla\cdot\left[{\bf b}\,({\bf
b}\cdot{\bf r})^{l-1}\right]=l(l-1)({\bf b}\cdot{\bf r})^{l-2}{\bf
b}\cdot{\bf b}=0. \label{ecu9a} \end{equation}

It can be shown that any $H_l$-harmonic can always be written as a
linear combination of the form
\begin{equation}
H_l({\bf r})= A_1 ({\bf b}_1 \cdot{\bf r})^l + A_2 ({\bf b}_2
\cdot{\bf r})^l + ... , \label{ecu9ab}
\end{equation}
for suitable constants $A_1, A_2, ...$ and null vectors ${\bf
b}_1, {\bf b}_2,... $ whence any property that can be easily found
out for the simplest solid harmonics can be extended by linearity
to any general $H_l$. To prove (\ref{ecu9ab}), let us consider the
particular set of null vectors of the form ${\bf b}=i\, \cos u
\,{\bf e}_x + i\, \sin u\, {\bf e}_y + {\bf e}_z$ $( 0 \le u < 2
\pi)$, or $({\bf b}\cdot{\bf r})^l=(i\,  x \cos u + i\,  y \sin u
+ z)^l$. On expressing the trigonometric functions in polar form
and binomially expanding we find for any given value $u_s$ in $(0,
2\pi)$ that
\begin{equation}
(i\,x \cos u_s  + i\, y \sin u_s + z)^l=\sum_{k'=-l}^{l}
{h_l}^{(k')} (x,y,z) e^{i k' u_s}, \label{ecu9ac}
\end{equation}
where the coefficients of the trigonometric series, i.e the
homogeneous polynomials ${h_l}^{(k')}(x,y,z)$, must constitute a
set $2l+1$ independent solid harmonics. By taking the set of
points $ u_s=2 \pi s/(2l+1)\,\,(s=0,...2l)$ and using the
orthogonality property $\sum_{s=0}^{2l} e^{-2\pi i s
(k-k')/(2l+1)}=(2l+1) \delta_{k k'} $ we can easily invert
(\ref{ecu9ac}) to obtain
\begin{equation}
 {h_l}^{(k)}
(x,y,z)=\sum_{s=0}^{2l} e^{2\pi i k s/(2l+1)} \left(i\, x \cos
\left[\frac{ 2\pi i k s}{2l+1}\right]   + i\, y \sin \left[\frac{
2\pi i k s}{2l+1}\right] + z\right)^l, \label{ecu9ad}
\end{equation}
whence each of the $2l+1$ independent solid harmonics ${h_l}^{(k)}
(x,y,z)$ can be expressed as a linear combination of solid
harmonics of the form $({\bf b}_j \cdot{\bf r})^l$. Since every
$H_l$-harmonic is a linear combination of the ${h_l}^{(k)}$'s,
equation (\ref{ecu9ab}) immediately follows from (\ref{ecu9ad}) .

An important application of (\ref{ecu9ab}) results from the
analysis of the repeated action of the differential operator ${\bf
b}\cdot \nabla$ - ${\bf b}$ being a constant null vector - on any
spherically symmetric function $F(r)$. In effect, since
\begin{equation}
{\bf b}\cdot \nabla F= {\bf b}\cdot {\bf e}_r \frac{dF}{dr}={\bf
b}\cdot {\bf r} \frac{1}{r}\frac{dF}{dr},\label{ecu9ad1}
\end{equation}
and
\begin{equation}
{\bf b}\cdot \nabla ({\bf b} \cdot{\bf r})^n=n ({\bf b} \cdot{\bf
r})^{n-1}{\bf b}\cdot \nabla ({\bf b} \cdot{\bf r})=n ({\bf b}
\cdot{\bf r})^{n-1} {\bf b}\cdot {\bf b}=0 \label{ecu9ad2}
\end{equation}
for any integer $n > 0$, on applying $l-1$ times ($l\ge 1$) the
operator ${\bf b}\cdot \nabla$ to (\ref{ecu9ad1}) we obtain
\begin{equation}
({\bf b}\cdot \nabla)^l F=({\bf b}\cdot {\bf r})^l
\left(\frac{1}{r}\frac{d}{dr}\right)^l F.\label{ecu9ad3}
\end{equation}
Equation (\ref{ecu9ab}) then implies
\begin{equation}
H_l(\nabla)F(r)=H_l({\bf r})
\left(\frac{1}{r}\frac{d}{dr}\right)^l F(r),\label{ecu9ad4}
\end{equation}
where $H_l(\nabla)$ is the differential operator obtained by
substituting each cartesian component of ${\bf r}$ in $H_l({\bf
r})$ by the corresponding one of $\nabla$.

As an illustration of (\ref{ecu9ad4}), let us consider the
integral (to be used in Section VI on partial wave expansions)
\begin{equation}
\int d\Omega_q H_{l}({\bf q}) e^{{\bf q}\cdot {\bf
r}}=H_{l}(\nabla) \int d\Omega_q e^{{\bf q}\cdot {\bf r}},
\label{ecpw1}
\end{equation}
where ${\bf q}=q_x{\bf e}_x + q_y{\bf e}_y + q_z {\bf e}_z$ is a
vector of the form and ${\bf q}=q {\bf e}_q$ - $q$ being a
(possibly complex) isotropic scalar and ${\bf e}_q$ a real unit
vector -, and $d\Omega_q\equiv \sin\theta_q d\theta_q d \varphi_q$
is the element of solid angle around the point ${\bf e}_q$ of the
unit sphere. The equality in (\ref{ecpw1}) easily follows from the
properties of the derivatives of the exponential function taking
into account that $H_{l}({\bf q})$ is a homogeneous polynomial.
Introducing $\mu_q=\cos \theta_q={\bf e}_q\cdot{\bf e}_r$, the
last integral in (\ref{ecpw1}) yields
\begin{equation}
\int d\Omega_q e^{{\bf q}\cdot{\bf r}} = 2 \pi \int_{-1}^{1}
d\mu_q e^{q r \mu_q}=4 \pi \frac{\sinh(qr)}{qr}, \label{ecpw2}
\end{equation}
so that, on using (\ref{ecu9ad4}),
\begin{equation}
\int d\Omega_q H_{l}({\bf q}) e^{{\bf q}\cdot {\bf r}}=4\pi
H_l({\bf r})  \left(\frac{1}{r}\frac{d}{dr}\right)^l
\left[\frac{\sinh(qr)}{qr}\right]. \label{120}
\end{equation}
 Defining the function
\begin{equation}
s_l(\zeta)\equiv \left(\frac{1}{\zeta}\frac{d}{d\zeta}\right)^l
\left(\frac{\sinh \zeta}{\zeta}\right)=\sum_{j=l}^{\infty}
\frac{2^l j!}{(j-l)!} \frac{ \zeta^{2(j-l)}}{(2j+1)!} \label{122b}
\end{equation}
we can write (\ref{120}) more briefly  as
\begin{equation}
\int d\Omega_q H_{l}({\bf q}) e^{{\bf q}\cdot {\bf r}}=4\pi
H_l({\bf r}) q^{2 l} s_l(qr).\label{120aa}
\end{equation}

Note that the series expansion for $s_l(\zeta)$ can be easily
obtained by applying the operator
$\left[(1/\zeta)d/d\zeta\right]^l=2^l\left[d/d\zeta^2\right]^l$ to
the power expansion of $\sinh \zeta/ \zeta$.  The fact that this
series only contains even powers of $\zeta$ makes possible to
compute (\ref{ecpw1}) for the case in which the vector ${\bf r}$
is replaced by the gradient operator $\nabla$. In effect, it is
easy to convince oneself that then the result must be formally the
same as (\ref{120}) but with $r=\sqrt{{\bf r}\cdot {\bf r}}$
replaced by $\sqrt{{\bf \nabla}\cdot {\bf
\nabla}}=\sqrt{\nabla^2}$:
\begin{equation}
\int d\Omega_q H_{l}({\bf q}) e^{{\bf q}\cdot {\bf \nabla}}= 4\pi
H_l({\bf r}) q^{2 l}
s_l\left(q\sqrt{\nabla^2}\right),\label{ecpw3}
\end{equation}
where the operator on the RHS must be computed by replacing
$\zeta^2$ by $(q\sqrt{\nabla^2})^2=q^2\nabla^2$ in the series
(\ref{122b}) which, therefore, yields only integral powers of the
operator $\nabla^2$.

\subsection {Maxwell's harmonics and Maxwell's integral theorem }
 Maxwell's harmonics \cite{Hob,Apple} are formed by taking $l$ directional derivatives of $1/r$
in any given directions defined by (generally complex) vectors
${\bf e}_1$,...${\bf e}_l$. Thus the function
\begin{equation}
V_l({\bf r})=({\bf e}_1\cdot \nabla)...({\bf e}_l\cdot
\nabla)(1/r)=\frac{\partial\,^l (1/r)}{\partial h_1...\partial
h_l}, \label{ecu20}
\end{equation}
where $\partial h_i$ $(i=1,...l)$ denotes the infinitesimal
(generally complex) displacement associated to ${\bf e}_i$, must
be a $V_l$-harmonic, since it can be seen by inspection that
$r^{2l+1} V_l$ is a homogeneous polynomial of degree $l$ which
satisfies Laplace's equation (since so does $V_l$). We will
illustrate this for the first few values of $l$, and then the
interested reader can easily proceed by induction\cite{Weeks} to
prove it for any $l$. For example, for $l=1$ and $l=2$ we have
\begin{equation}
{\bf e}_1\cdot \nabla \left(1/r \right) =-\frac{{\bf e}_1\cdot{\bf
r}}{r^3} \label{ecu21}
\end{equation}
and
\begin{equation}
{\bf e}_1\cdot \nabla \left[{\bf e}_2\cdot \nabla \left(1/r
\right)\right] =-\frac{3\,{\bf e}_1\cdot{\bf r}\,\,{\bf
e}_2\cdot{\bf r}-r^2\,{\bf e}_1\cdot{\bf e}_2}{r^5}, \label{ecu22}
\end{equation}
where it can be readily checked that the homogeneous polynomials
in the numerators satisfy Laplace's equation.

Therefore, the function
\begin{equation}
H_l({\bf r})=r^{2l+1}\frac{\partial\,^l (1/r)}{\partial
h_1...\partial h_l} \label{ecu22a}
\end{equation}
is an $H_l$-harmonic, and the function
\begin{equation}
Y_l({\bf r})=r^{l+1} \frac{\partial\,^l (1/r)}{\partial
h_1...\partial h_l} \label{ecu23}
\end{equation}
is a surface harmonic of degree $l$. Solid harmonics which, except
for a constant factor, are of the form (\ref{ecu20}) and
(\ref{ecu22a}) will be called {\it Maxwell's solid harmonics} of
the $V_l$ and $H_l$ types, respectively, and surface harmonics of
the form (\ref{ecu23}) will be called {\it Maxwell's surface
harmonics}. The vectors ${\bf e}_i$ are called the {\it poles} of
the Maxwell's harmonics, and in the particular case in which they
are both unit and real, they define $l$ points on the unit sphere
known as {\it Maxwell's poles}\cite{Hob,Apple,Dowker,Weeks}.

A Maxwell's surface harmonic in which all the ${\bf e}_i$'s
coincide with a given vector, say ${\bf e}_1$, is called a {\it
zonal harmonic}:
\begin{equation}
Z_l({\bf r})=r^{l+1} {({\bf e}_1\cdot
\nabla)}^l\left(1/r\right)=r^{l+1} \frac{\partial\,^l
(1/r)}{\partial {h_1}^l} \label{ecu23aa}
\end{equation}
Zonal harmonics arise most naturally when one Taylor expands
around ${\bf r=0}$ the potential due to a unit source localized at
${\bf r'}$. For instance, the expansion valid for $r
> r'$ can be written as
\begin{equation}
\frac{1}{|{\bf r}-{\bf r'}|}= \sum_{l=0}^{\infty}
\frac{r'^l}{r^{l+1}} \frac{(-1)^{l} r^{l+1} }{l!} \left({\bf
e}_r'\cdot \nabla\right)^l\left(1/r\right), \label{ecu23ab}
\end{equation}
where ${\bf r'}=r'{\bf e}_r'$. Note that all factors
$r'^l/r^{l+1}$ in the series (\ref{ecu23ab}) are multiplied by a
zonal harmonic of the form (\ref{ecu23aa}) with pole at ${\bf
e}_r'$. On the other hand, the well known\cite{Jack} expansion in
terms of Legendre's polynomials, $P_l$, is
\begin{equation}
\frac{1}{|{\bf r}-{\bf r'}|}=\sum_{l=0}^{\infty}
\frac{r'^l}{r^{l+1}} P_l({\bf e}_r'\cdot{\bf e}_r),
\label{ecu23ac}
\end{equation}
whence identifying terms in (\ref{ecu23ab})-(\ref{ecu23ac}) we
obtain:
\begin{equation}
P_l({\bf e}_r'\cdot{\bf e}_r)=\frac{(-1)^{l} r^{l+1}}{l!}
\left({\bf e}_r'\cdot \nabla\right)^{l}\left(1/r\right),
\label{ecu23ad}
\end{equation}
which establishes the relation between Legendre polynomials and
zonal harmonics.

{\it Maxwell's integral theorem} results when we substitute $Y_l$
in (\ref{ecu13A4}) by a surface harmonic of the form
(\ref{ecu23}). Then $V_l=r^{-(l+1)} Y_l$ is given by (\ref{ecu20})
and, since $\nabla^2(1/r)=-4\pi\delta({\bf r})$ - where
$\delta({\bf r})$ is Dirac's delta function - we have
\begin{equation}
 \nabla^2 V_l({\bf r})=-4\pi \frac{\partial\,^l  \delta({\bf
r})}{\partial h_1...\partial h_l}. \label{ecu23adb}
\end{equation}
Substituting (\ref{ecu23adb}) into the last integral of
(\ref{ecu13A4}) and integrating $l$ times by parts - taking into
account that $\delta({\bf r})$ and all its partial derivatives
vanish at the surface $r=a$ -, we obtain \begin{equation}
 \int_{r=a} dS \Phi Y_l =  \frac{4\pi (-1)^{l} a^{l+2}}{2l+1}
\frac{\partial\,^l \Phi }{\partial h_1...\partial h_l}\Big
|_{r=0}. \label{ecu13adc}
\end{equation}
which is Maxwell's integral theorem. The direct proof of
(\ref{ecu13adc}) given here is the only one we are aware of
besides that given by Maxwell himself in his treatise. It seems to
us than ours is somewhat more transparent than his - which is
based on the properties singular multipolar charge distributions
-.

As an illustration, let us apply (\ref{ecu13adc}) for the case in
which $\Phi$ is any solid harmonic $H_l$ and the Maxwell's
harmonic is the zonal harmonic $Z_l$ given by (\ref{ecu23aa}):
\begin{equation} \int_{r=a} dS H_l Z_l = \frac{4\pi (-1)^{l}
a^{l+2}}{2l+1} \left[({\bf e}_1 \cdot\nabla)^l H_l\right]_{r=0}.
\label{ecu23k}
\end{equation}
In order to compute the RHS of (\ref{ecu23k}) we  binomially
expand the operator $({\bf e}_1\cdot\nabla)^l$,
\begin{equation}
({\bf e}_1\cdot\nabla)^l = l! \sum_{\alpha'+\beta'+\gamma'=l}
\frac{e_{1x}^{\alpha'}e_{1y}^{\beta'}e_{1z}^{\gamma'}}{\alpha'! \,
\beta'! \,\gamma'!} \frac{\partial^{\alpha'}}{\partial
x^{\alpha'}} \frac{\partial^{\beta'}}{\partial
y^{\beta'}}\frac{\partial^{\gamma'}}{\partial
z^{\gamma'}},\label{ecu23jjjj}
\end{equation}
and take into account the expression (\ref{ecu6}) for $H_l({\bf
r})$. On computing the result at ${\bf r}=0$ we obtain
\begin{equation}
\left[({\bf e}_1\cdot\nabla)^l H_l({\bf r})\right]_{r=0} = l!
\sum_{\alpha+\beta+\gamma=l} C_{\alpha \beta \gamma}
e_{1x}^{\alpha}e_{1y}^{\beta} e_{1z}^{\gamma}=l! H_l({\bf
e}_1),\label{ecu23j}
\end{equation}
which is a direct consequence of the homogeneity of $H_l$ - and
therefore it remains valid if ${\bf e}_1$ and $H_l({\bf r})$ are
replaced by any constant (generally complex) vector and any
homogeneous polynomial of degree $l$, respectively-. Thus
\begin{equation} \int_{r=a} dS H_l Z_l = \frac{4\pi (-1)^{l} l! a^{l+2}}{2l+1} H_l({\bf e}_1).
\label{ecu23ka}
\end{equation}

 \section{The standard set of spherical harmonics $Y_{lm}$.}

The standard {\it spherical harmonics} \cite{Jack} of degree $l$,
$Y_{lm}({\bf r})$ $(m=-l,...l)$, are $2l + 1$ independent,
normalized surface harmonics such that the $m$-th one depends on
the azimuthal coordinate as $e^{im\varphi}$. The normalization
condition is defined as
\begin{equation}
\int d\Omega  Y^*_{lm'} Y_{lm} =\delta_{m'm}. \label{ecu26aaa}
\end{equation}
As already pointed out, these functions are usually given in terms
of the associated Legendre functions, $P_{l}^{m} (\cos\theta)$,
which can be introduced by separating variables in Laplace's
equation expressed in spherical coordinates (see also Appendix A).
However, we next show how they can be generated in an almost
trivial fashion by the procedure due to Thomson and Tait\cite{TT},
who constructed Maxwell's harmonics by taking directional
derivatives of $1/r$ with respect to $z$ and with respect to the
complex coordinates
\begin{equation}
\xi=x+iy=r \cos\theta e^{i  \varphi}\quad {\rm and}\quad
\eta=x-iy=r \cos\theta e^{-i  \varphi}. \label{ecu23add}
\end{equation}
In effect, let us assume for the moment that $m\ge 0$ and consider
the $V_l$-harmonic
\begin{equation}
V_{lm}({\bf r})\equiv \frac{\partial ^{l} (1/r)}{\partial z^{l-m}
\partial \eta^m}=(-1/2)^m\,(2m-1)!!\,
\xi^m \, \frac{\partial ^{l-m}}{\partial z^{l-m}}
\frac{1}{r^{2m+1}}, \label{ecu31}
\end{equation}
where we have carried out  $mth$-partial derivative with respect
to $\eta$ of $r^{-1}=(\eta\xi+z^2)^{-1/2}$ using the notation
$(2m-1)!!\equiv (2m-1)\times(2m-3)\times...1$ with the convention
$(-1)!!\equiv 1$. Observe that the presence of the factor $\xi^m$
automatically yields an azimuthal dependence of the form
$e^{im\varphi}$ in (\ref{ecu31}) . Therefore, the spherical
harmonics $Y_{lm}$ can be obtained from the $V_{lm}$'s by just
writing
\begin{equation}
Y_{lm}=C_{lm} r^{l+1} V_{lm}, \label{ecu38}
\end{equation}
where the constants $C_{lm}$ must be determined from the
normalization condition.  We can readily compute the normalization
integral (\ref{ecu26aaa}) via Maxwell's integral theorem by
putting $\Phi\equiv r^l Y^*_{lm}$, $Y_l\equiv Y_{lm}$ and $a=1$ in
(\ref{ecu13adc}), which yields
\begin{equation} \frac{4\pi (-1)^{l} |C_{lm}|^2}{2l+1}
\frac{\partial^l (r^{2l+1} V^*_{lm})}{\partial z^{l-m}
\partial \eta^m} \Big|_{r=0}=1.
\label{ecu40} \end{equation} Note that, since $r^{2l+1} V^*_{lm}$
is a homogeneous polynomial of degree $l$ in the variables $\xi$,
$\eta$ and $z$, only the coefficient of the term  containing
$\eta^m z^{l-m}$ contributes to the value at $r=0$ of the
derivative in (\ref{ecu40}). Multiplying (\ref{ecu31}) by
$r^{2l+1}$ and taking complex conjugates we obtain
\begin{equation}
r^{2l+1} V^*_{lm}=(-1/2)^m\,(2m-1)!!\,\eta^m \, r^{2l+1}
\frac{\partial ^{l-m}}{\partial z^{l-m}} \frac{1}{r^{2m+1}},
\label{ecu40a}
\end{equation}
so that in order to find the coefficient of $\eta^m z^{l-m}$ in
(\ref{ecu40a}) we just have to compute the value at $\eta=0$ and
$\xi=0$ of the factor to the right of $\eta^m$ , namely
\begin{equation}
z^{2l+1} \frac{d ^{l-m} }{d z^{l-m}}\frac{1}{z^{2m+1}}= (-1)^{l-m}
\frac{(l+m)!}{(2m)!} z^{l-m}. \label{ecu40aa}
\end{equation}
Substituting (\ref{ecu40aa}) into (\ref{ecu40a}) we  obtain
\begin{equation} r^{2l+1}V^*_{lm}=\frac{(-1)^l\,(2m-1)!!(l+m)!}{2^m (2m)!}  \eta^m
z^{l-m} + ..., \label{ecu40ab}
\end{equation}
where the dots denote terms which do not contribute to the LHS of
(\ref{ecu40}), whence
\begin{equation} \frac{4\pi (l+m)!(l-m)!|C_{lm}|^2}{(2l+1)\,
2^{2m}}=1 . \label{ecu40abc}
\end{equation}
The normalization constant $C_{lm}$ is determined by
(\ref{ecu40abc}) except for a phase factor which we will choose as
$(-1)^{l+m}$. This choice is motivated so that the resulting
spherical harmonics satisfy the relation (\ref{ecu48aad}) below,
as do those used in standard references \cite{Jack,Sch}. Solving
(\ref{ecu40abc}) for $C_{lm}$ and substituting into (\ref{ecu38})
we finally obtain
\begin{equation}
Y_{lm}({\bf r})= \sqrt{\frac{2l+1}{4\pi}} \frac{\,(-1)^{l+m}
2^{m}\,r^{l+1} } {\sqrt{(l+m)!(l-m)!}} \frac{\partial ^{l}
(1/r)}{\partial z^{l-m} \partial \eta^m}. \label{ecu45}
\end{equation}
It is shown in Appendix A that (\ref{ecu45}) reproduces the
standard expression\cite{Luca} for the $Y_{lm}$'s when expressed
in spherical coordinates.

Until now we have assumed  $0 \le m \le l$ and found a set of
$l+1$  spherical harmonics given by (\ref{ecu45}) for
$m=0,1,...l$. These functions, together with their corresponding
complex conjugates $Y^*_{lm}({\bf r})$ form a complete set of
$2l+1$ surface harmonics
$\big{\{}\,Y_{l0},\,Y_{lm},\,Y^*_{lm}\quad m=1,2...l\big{\}}$.
However, in actual calculations it is often more convenient to
work with a set of $2l+1$ functions labelled by an index $m$
running from $-l$ to $l$, and to arrange things so that, for
negative $m$, $Y_{lm}$ is also given by an expression of the form
(\ref{ecu45}) conveniently defined for $m<0$. For this purpose, we
first note the identity
\begin{equation}
\nabla^2 (1/r)=4 \frac{\partial^2 (1/r)}{\partial \eta \partial
\xi}+\frac{\partial^2 (1/r)}{\partial z^2}=0 \label{ecu48aa}
\end{equation}
obtained after expressing the Laplacian operator in coordinates
($\xi, \eta, z$) through the chain rule. The result
(\ref{ecu48aa}) suggests that we can give an operational meaning
to symbols such as $\left(\partial/\partial \eta\right)^{-1}$ and
$\left(\partial/\partial z\right)^{-2}$ {\it when applied to
$(1/r)$} by means of the relations
\begin{equation}
4 \frac{\partial^2 }{\partial \eta \partial \xi}=-\frac{\partial^2
}{\partial z^2}\rightarrow 4 \frac{\partial}{\partial
\xi}=-\left(\frac{\partial}{\partial
\eta}\right)^{-1}\frac{\partial^2}{\partial z^2}\rightarrow 4
\left(\frac{\partial} {\partial
z}\right)^{-2}\frac{\partial}{\partial
\xi}=-\left(\frac{\partial}{\partial \eta}\right)^{-1},
\label{ecu48aab}
\end{equation}
which allow us to define $V_{lm}$ for $m < 0$ in terms of
$V^*_{l|m|}$ by means of (\ref{ecu31}) . In effect, according to
(\ref{ecu31}) we can formally write
\begin{equation}
 V_{l,-|m|}\equiv \left(\frac{\partial}{\partial z}\right)^{l+|m|}\left(\frac{\partial }
{\partial \eta}\right)^{-|m|} \left(\frac{1}{r}\right)
\label{ecu48aab1}
\end{equation}
and, if we substitute $\left(\partial/\partial \eta\right)^{-|m|}$
by the  $|m|$th-power of the RHS of the last operator equation in
(\ref{ecu48aab}) we obtain
\begin{equation}
 V_{l,-|m|}=(-4)^{|m|}\frac{\partial ^{l} (1/r)}{\partial z^{l-|m|}
\partial \xi^{|m|}}=(-4)^{|m|} V^*_{l|m|}.
\label{ecu48aab2}
\end{equation}
But it is easily seen that if we drop the absolute value sign in
(\ref{ecu48aab2}) the resulting equation,
\begin{equation}
 V_{l,-m}= (-1)^m 4^{m} V^*_{lm},
\label{ecu48aac}
\end{equation}
remains consistent with definition (\ref{ecu48aab2}) independently
of the sign of $m$, whence (\ref{ecu48aac}) can be freely used to
relate the $V_{lm}$'s  of positive and negative $m$. This result
will be of great use in Section IV on recursion relations for
spherical harmonics. The corresponding relation for the $Y_{lm}$'s
can be now obtained from (\ref{ecu38}) after noticing that
replacing $m$ by $-m$ in the normalization constant in
(\ref{ecu45}) yields the new constant $C_{l,-m}=C_{lm}/4^m$, or
\begin{equation}
 Y_{l,-m}= C_{l,-m} r^{l+1} V_{l,-m}= (-1)^m
 Y^*_{lm}.
\label{ecu48aad}
\end{equation}

The functions $Y_{lm}$ with $-l\le m \le l$ defined by
(\ref{ecu45}) and (\ref{ecu48aad}) constitute the set of standard
spherical harmonics of degree $l$.

\section{Recursion relations.}
 Recursion relations are of importance in calculations involving
spherical harmonics,  a well known application in Quantum
Mechanics being the computation of the matrix elements of
operators in the angular momentum representation \cite{Kra}. In
classical physics the best known examples are perhaps the analysis
of multipolar radiation fields \cite{Jack,Sch} and wave scattering
problems solved numerically using the relatively recent {\it Fast
Multipole Method}\cite{Rokhlin, Rusos}. The standard derivations
of such recursion relations, i.e., using the associated Legendre
functions and spherical polar coordinates, often get rather tricky
and involved, yielding messy final expressions\cite{But,Abra}. In
this section we indicate an apparently new procedure to
systematically derive a whole set of useful recursion relations
using the Maxwell solid harmonics $V_{lm}({\bf r})$ defined in
Section III [see (\ref{ecu31})].

We begin by noticing the equality
\begin{equation}
z\frac{\partial ^{l-m}}{\partial z^{l-m}}\frac{\partial
^m}{\partial \eta^m}\left(\frac{1}{r}\right)=\frac{\partial
^{l-m}}{\partial z^{l-m}}\frac{\partial ^m}{\partial
\eta^m}\left(\frac{z}{r}\right)-(l-m) \frac{\partial
^{l-m-1}}{\partial z^{l-m-1}}\frac{\partial ^m}{\partial
\eta^m}\left(\frac{1}{r}\right),\label{ecu53}
\end{equation}
where we have applied Leibnitz's rule to the derivatives with
respect to $z$ of $(z/r)$. Since $r=\sqrt{\xi \eta + z^2}$,
substituting the relation
\begin{equation}
\frac{\partial}{\partial \eta} \left(\frac{z}{r}\right)=-\frac{\xi
z}{2r^{3/2}}=\frac{\xi}{2} \frac{\partial}{\partial z}
\left(\frac{1}{r}\right) \label{ecu54}
\end{equation}
into the first summand on the RHS of (\ref{ecu53}) we obtain
\begin{equation}
z\frac{\partial ^{l-m}}{\partial z^{l-m}}\frac{\partial
^m}{\partial \eta^m}\left(\frac{1}{r}\right)=\frac{\xi}{2}
\frac{\partial ^{l-m+1}}{\partial z^{l-m+1}}\frac{\partial
^{m-1}}{\partial \eta^{m-1}}\left(\frac{1}{r}\right)-(l-m)
\frac{\partial ^{l-m-1}}{\partial z^{l-m-1}}\frac{\partial
^m}{\partial \eta^m}\left(\frac{1}{r}\right)\label{ecu54a}
\end{equation}
or, in terms of the $V_{lm}$'s defined in (\ref{ecu31}),
\begin{equation}
2z V_{lm}=\xi V_{l,m-1}-2(l-m)V_{l-1,m},\label{ecu56}
\end{equation}
which is our first recursion relation. A corresponding one for
$\eta V_{lm}$ can be obtained by replacing $m$ by $-m$ in
(\ref{ecu56}) and taking complex conjugates using
(\ref{ecu48aac}). This yields, after replacing $m$ by $m-1$ in the
resulting equation,
\begin{equation}
2\eta V_{lm}=-z V_{l,m-1}-(l+m-1)V_{l-1,m-1}.\label{ecu61a}
\end{equation}
Also, adding (\ref{ecu56}) multiplied by $z$ to (\ref{ecu61a})
multiplied by $\xi$ we find the relation
\begin{equation}
2r^2 V_{lm}=-2(l-m)z V _{l-1,m}-(l+m-1)\xi V
_{l-1,m-1}.\label{ecu61}
\end{equation}
Replacing $l$ by $l+1$ in (\ref{ecu61}), substituting $z V_{lm}$
from  and, finally, replacing $m$ by $m+1$ we obtain
\begin{equation}
(2l+1)\xi V_{lm}=-2r^2 V_{l+1,m+1} + 2(l-m)(l-m-1) V_{l-1,m+1}.
\label{ecu62}
\end{equation}
Observe that  the factor $\xi$ multiplying $V_{lm}$ in
(\ref{ecu62}) depends on the angular coordinates $\theta$ and
$\varphi$ while none of the factors multiplying the functions $V$
on the RHS does so. Therefore, recursion relations such as
(\ref{ecu62}) are very convenient to compute surface integrals
(projections) of the product of the function appearing in the LHS
times  Maxwell's solid or surface harmonics, since then we can use
the orthogonality properties of the latter with those on the RHS.
Analogous expression for $\eta V_{lm}$ can be obtained by
replacing $m$ by $-m$ in (\ref{ecu62}) and taking complex
conjugates using (\ref{ecu48aac}), which yields
\begin{equation}
2(2l+1)\eta V _{lm}=r^2 V _{l+1,m-1}-(l+m)(l+m-1) V
_{l-1,m-1},\label{ecu63}
\end{equation}
The corresponding expression for $z V_{lm}$, is obtained  by
replacing $m$ by $m-1$ in (\ref{ecu62}) and substituting into
(\ref{ecu56}):
\begin{equation}
(2l+1)z V _{lm}=-r^2 V_{l+1,m}-(l-m)(l+m) V_{l-1,m},\label{ecu64}
\end{equation}

In some circumstances, it is convenient to express a derivative
with respect to either $z$, $\eta$ or $\xi$ of the product of
$V_{lm}$ times a function of $r$, say $B(r)$, in terms of
functions $V_{lm}$ multiplied by coefficients which only depend on
$r$. This is the case, for example, in the computation of matrix
elements of the quantum momentum operator in the angular momentum
representation\cite{Kra}. Since $r=\sqrt{\xi\eta + z^2}$, applying
the chain rule we obtain
\begin{equation}
\frac{\partial}{\partial z}\left(B V_{lm}\right)=B'z V_{lm}/r + B
V _{l+1,m},\label{ecu65}
\end{equation}
\begin{equation}
\frac{\partial}{\partial \eta}\left(B V _{lm}\right)=B'\xi V
_{lm}/(2r) + B V_{l+1,m+1},\label{ecu66}
\end{equation}
\begin{equation}
\frac{\partial}{\partial \xi}\left(B V _{lm}\right)=B'\eta V
_{lm}/(2r) - B V_{l+1,m-1}/4,\label{ecu67}
\end{equation}
where we have written $B'\equiv(dB/dr)$ and used the relations
\begin{equation}
\frac{\partial V_{lm}}{\partial z}=\frac{\partial
^{l-m+1}}{\partial z^{l-m+1}}\frac{\partial ^{m}}{\partial
\eta^{m}}\left(\frac{1}{r}\right)=V_{l+1,m},\quad\frac{\partial
V_{lm}}{\partial \eta}=\frac{\partial ^{l+1-(m+1)}}{\partial
z^{l+1-(m+1)}}\frac{\partial ^{m+1}}{\partial
\eta^{m+1}}\left(\frac{1}{r}\right)=V_{l+1,m+1}.\label{ecu54b}
\end{equation}
Using the recursion relations (\ref{ecu62})-(\ref{ecu64}) in
(\ref{ecu65})-(\ref{ecu67}) we readily obtain the sought
expressions.

It can be checked that the recursion relations here obtained
reproduce those derived in a different way by Kramers\cite{Kra}
(Ch.IV p. 178). For this, it is necessary to replace in all of
them $l$ by $l-1$, to make $B(r)\equiv r^{l}b(r)$, and to express
them in terms of the (unnormalized) surface harmonics
$\mathcal{P}_{lm}$ used by Kramers:
\begin{equation}
\mathcal{P}_{lm}=(-1)^l \frac {2^m l!}{(l-m)!(l+m)!} r^{l+1}
V_{lm}.\label{ecu71}
\end{equation}
 Nevertheless, it seems to us that our derivations and
expressions are simpler and more systematic than Kramer's due to
we use solid harmonics, the $V_{lm}$'s, instead of surface
harmonics.

Finally, of special interest in Quantum Mechanics are the ladder
properties of the operator ${\bf r}\times \nabla$-which is
proportional to the angular momentum operator- when applied to the
spherical harmonics $Y_{lm}$. Notice that, since ${\bf r}\times
\nabla r^{-(l+1)}=0$, its effect on the $Y_{lm}$'s will be the
same as that on the functions $V_{lm}$. In order to analyze the
operator ${\bf r}\times \nabla$ in a coordinate-free manner, it is
convenient at this point to introduce the so-called {\it spherical
basis} formed by the vector ${\bf e}_z$ together with the vectors
\begin{equation}
{\bf e}_\xi=\left({\bf e}_x-i{\bf e}_y\right)/\sqrt{2}\quad{\rm
and}\quad{\bf e}_\eta=\left({\bf e}_x+i{\bf
e}_y\right)/\sqrt{2}={\bf e}^*_\xi.\label{ecu72}
\end{equation}
It is immediately seen that these vectors satisfy the
orthogonality relations
\begin{equation}
{\bf e}_\xi\cdot {\bf e}_\xi={\bf e}_\eta\cdot {\bf e}_\eta=0\quad
{\rm and}\quad {\bf e}_\xi\cdot {\bf e}_\eta=1 \label{ecu72aaa}
\end{equation}
and
\begin{equation}
{\bf e}_\xi\times{\bf e}_\eta=i{\bf e}_z,\quad{\bf
e}_\eta\times{\bf e}_z=i{\bf e}_\eta,\quad {\bf e}_z\times{\bf
e}_\xi=i{\bf e}_\xi.\label{ecu72aaa1}
\end{equation}
Note that (\ref{ecu72aaa}) imply that $ {\bf e}_\xi $ and $ {\bf
e}_\eta$ are null vectors. In terms of the spherical vectors, the
cartesian vectors $ {\bf e}_x $ and $ {\bf e}_y$ are given by
\begin{equation}
{\bf e}_x=\left({\bf e}_\xi+{\bf e}_\eta\right)/\sqrt{2}\quad{\rm
and}\quad{\bf e}_y=i\,\left({\bf e}_\xi-{\bf
e}_\eta\right)/\sqrt{2},\label{ecu72aa}
\end{equation}
so that any vector ${\bf u}$ can be written as
\begin{equation}
{\bf u}=u_x {\bf e}_x + u_y {\bf e}_y + u_z {\bf e}_z=u_\xi {\bf
e}_\xi + u_\eta {\bf e}_\eta + u_z {\bf e}_z,\label{ecu72aa1}
\end{equation}
with the components $u_\xi$ and $u_\eta$ given by
\begin{equation}
{\bf u}\cdot {\bf e}_\eta=u_\xi=(u_x+i u_y)/\sqrt{2}\quad{\rm
and}\quad {\bf u}\cdot {\bf e}_\xi=u_\eta=(u_x-i
u_y)/\sqrt{2}.\label{ecu72aa2}
\end{equation}
In particular, for the position vector we have
\begin{equation}
{\bf r}=\xi {\bf e}_\xi /\sqrt{2} + \eta {\bf e}_\eta /\sqrt{2} +
z {\bf e}_z,   \label{ecu72aa21}
\end{equation}
and  the gradient operator is
\begin{equation}
{\bf \nabla}=  {\bf e}_x \frac{\partial}{\partial x} +  {\bf e}_y
\frac{\partial}{\partial y}
 + {\bf e}_z \frac{\partial}{\partial z}= \sqrt{2} {\bf e}_\xi \frac{\partial}{\partial \eta}
+  \sqrt{2} {\bf e}_\eta \frac{\partial}{\partial \xi}
 + {\bf e}_z \frac{\partial}{\partial z},\label{ecu72aa21a}
\end{equation}
where we have used (\ref{ecu72aa}) together with the relations
between the partial derivatives with respect to the cartesian and
the complex coordinates - obtained via the chain rule using
(\ref{ecu23add}) - :
\begin{equation}
 \frac{\partial}{\partial x} =\frac{\partial}{\partial \xi}+\frac{\partial}{\partial
 \eta}\quad{\rm and}\quad\frac{\partial}{\partial y} =i\frac{\partial}{\partial \xi}-i\frac{\partial}{\partial
 \eta}. \label{ecu72aa21b}
\end{equation}

 Using the recursion relations and the properties of the spherical basis derived above,
 it is easy to show that the operator ${\bf e}_\eta\cdot({\bf r}\times \nabla)$
rises the index $m$ of functions $V_{lm}$ $(m=-l,-l+1,...l-1,l)$
on which it acts, while the operator ${\bf e}_\xi\cdot({\bf
r}\times \nabla)$ lowers it. In effect, taking into account that
\begin{equation}
{\bf e}_\eta\times{\bf r}=\xi{\bf e}_\eta\times{\bf
e}_\xi/\sqrt{2}+z {\bf e}_\eta\times{\bf e}_z=-i\xi{\bf
e}_z/\sqrt{2}+iz{\bf e}_\eta\label{ecu71a}
\end{equation}
and that ${\bf e}_\eta\cdot\nabla=\sqrt{2}\partial/\partial \eta$
[see (\ref{ecu72aa21a})], we find
\begin{equation}
{\bf e}_\eta\cdot({\bf r}\times \nabla V_{lm})=({\bf
e}_\eta\times{\bf r})\cdot \nabla V_{lm}=-i\xi/\sqrt{2}
V_{l+1,m}+\sqrt{2} i z V_{l+1,m+1} ,\label{ecu76666}
\end{equation}
or, using recursion relation (\ref{ecu56}),
\begin{equation}
{\bf e}_\eta\cdot({\bf r}\times \nabla
V_{lm})=-\sqrt{2}i(l-m)V_{l,m+1}.\label{ecu76}
\end{equation}
The corresponding property for the operator ${\bf e}_\xi\cdot({\bf
r}\times \nabla)$ can be obtained by replacing $m$ by $-m$ in
(\ref{ecu76}) and taking the complex conjugate of the resulting
expression. Thus relations  (\ref{ecu48aac}) and (\ref{ecu56})
lead to
\begin{equation}
{\bf e}_\xi\cdot({\bf r}\times \nabla
V_{lm})=-i(l+m)V_{l,m-1}/\sqrt{2}. \label{ecu77}
\end{equation}
In terms of the spherical harmonics (\ref{ecu38})  equations
(\ref{ecu76}) and (\ref{ecu77}) read
\begin{equation}
{\bf e}_\eta\cdot({\bf r}\times \nabla
Y_{lm})=i\sqrt{(l-m)(l+m+1)}Y_{l,m+1}/\sqrt{2}.\label{ecu78}
\end{equation}
\begin{equation}
{\bf e}_\xi\cdot({\bf r}\times\nabla
Y_{lm})=i\sqrt{(l+m)(l-m+1)}Y_{l,m-1}/\sqrt{2}.\label{ecu79}
\end{equation}

Relations (\ref{ecu78}) and (\ref{ecu79}) are usually expressed in
terms of spherical polar coordinates\cite{Sch}, $\theta$ and
$\varphi$. To do this, we first obtain the cartesian components of
the operator
\begin{equation}
{\bf r}\times\nabla ={\bf e}_\varphi \frac{\partial}{\partial
\theta}- \frac{{\bf
e}_\theta}{\sin{\theta}}\frac{\partial}{\partial \varphi}
\label{ecu80}
\end{equation}
by recalling that ${\bf e}_\varphi= -\sin\varphi{\bf e}_x
+\cos\varphi{\bf e}_y$ and ${\bf e}_\theta={\bf
e}_\varphi\times{\bf e}_r$ which, in turn, imply ${\bf
e}_x\cdot{\bf e}_\theta=({\bf e}_x\times{\bf e}_\varphi)\cdot{\bf
e}_r=\cos{\varphi}\cos\theta$ and ${\bf e}_y\cdot{\bf
e}_\theta=\sin{\varphi}\cos\theta$. Thus
\begin{equation}
{\bf r}\times\nabla =-\left({\bf e}_x \sin\varphi-{\bf e}_y
\cos\varphi\right)\frac{\partial}{\partial \theta}-\left({\bf e}_x
\cos\varphi+{\bf e}_y \sin\varphi\right)
\cot\theta\frac{\partial}{\partial \varphi}, \label{ecu81}
\end{equation}
and taking into account that ${\bf e}_\eta\cdot{\bf e}_x=2^{-1/2}$
and ${\bf e}_\eta\cdot{\bf e}_y=i 2^{-1/2}$
(\ref{ecu78})-(\ref{ecu79}) can be written in compact form as
\begin{equation}
e^{\pm im\varphi}\left(\pm\frac{\partial}{\partial
\theta}+i\cot\theta\frac{\partial}{\partial
\varphi}\right)Y_{lm}=\sqrt{(l\mp m)(l\pm m+1)}Y_{l,m\pm1}.
\label{ecu82}
\end{equation}

\section{Legendre's addition theorem.}
Maxwell's integral theorem leads to a very simple derivation of
the well known - and physically very useful - Legendre's addition
theorem\cite{Wenig2}. In effect, if we  restrict ${\bf e_1}$ in
(\ref{ecu23k}) to be a real, unit vector then $H_l({\bf e}_1)=
Y_l({\bf e}_1)$ - where  $Y_l\equiv H_l/r^l$- and, after dividing
(\ref{ecu23k}) by $a^l$, we obtain
\begin{equation}
\int_{r=a} dS Y_l Z_l = \frac{4\pi (-1)^{l} l! a^{2}}{2l+1}
Y_l({\bf e}_1), \label{ecu23l}
\end{equation}
which must hold for any surface harmonic $Y_l$. Therefore, if we
take any set of $2l+1$ orthonormal surface harmonics of degree
$l$, $\big{\{}\,X_{ln}({\bf r})\,:\,\int d\Omega X_{ln} X^*_{ln'}
=\delta_{nn'} \,\, (n=1,...2l+1)\big{\}}$ , the coefficients in
the expansion
\begin{equation}
Z_l({\bf r})=\sum_{n=1}^{2l+1} A_n X_{ln}({\bf r})
\label{ecu23mmm}
\end{equation}
are given by
\begin{equation}
A_n=\int d\Omega X^*_{ln} Z_l = \frac{4\pi (-1)^{l} l!}{2l+1}
X^*_{ln}({\bf e}_1), \label{ecu23n}
\end{equation}
where we have used (\ref{ecu23l}) with $a=1$. Thus
\begin{equation}
Z_l({\bf r})=\frac{4\pi (-1)^{l}  l!}{2l+1} \sum_{n=1}^{2l+1}
X^*_{ln}({\bf e}_1) X_{ln}({\bf r}), \label{ecu23o}
\end{equation}
and using the relation (\ref{ecu23ad}) to express $Z_l= (-1)^{l}
l! P_l({\bf e}_1\cdot{\bf e}_r)$ we finally obtain the celebrated
{\it Legendre's addition theorem}:
\begin{equation}
P_l({\bf e}_1\cdot{\bf e}_r)=\frac{4\pi}{2l+1} \sum_{n=1}^{2l+1}
X^*_{ln}({\bf e}_1) X_{ln}({\bf e}_r), \label{ecu23p}
\end{equation}
where we have taken into account that $X_{ln}({\bf r})=X_{ln}({\bf
e}_r)$ since the $X_{ln}$'s are surface harmonics. Notice that
(\ref{ecu23p}) reproduces the familiar form of the addition
theorem \cite{But, Jack} for the particular case in which the set
$X_{ln}$ $(n=0,...2l+1)$ is chosen to be the set of standard
spherical harmonics, $Y_{lm}({\bf r})$ $(m=-l,..,0,...+l)$,
introduced in the Section III.

\section{ Surface harmonics expansions and the completeness relation.}
Let us consider an arbitrarily thin sphere of radius $a$ charged
with a given surface charge density distribution
$f(\theta,\varphi)$. If ${\bf r'}=a {\bf e}_r'\,$ denotes the
generic point on the sphere, the electric potential at any point
${\bf r}$ of space is
\begin{equation}
\phi({\bf r})=\int_{r'=a} d S' \frac{f(\theta',\varphi')}{|{\bf
r}-{\bf r'}|}. \label{ecu23ac1}
\end{equation}
Inserting the expansion (\ref{ecu23ac}) into (\ref{ecu23ac1}) we
obtain
\begin{equation}
\phi({\bf r})= \sum_{l=0}^{\infty} \frac{a^l}{r^{l+1}} \int_{r'=a}
d S' P_l({\bf e}_r'\cdot{\bf e}_r) f(\theta',\varphi')
\label{ecu23ac2}
\end{equation}
for $r > a$ and, on using the expansion analogous to
(\ref{ecu23ac}) but with the roles of ${\bf r}$ and ${\bf r}'$
interchanged - i.e., the expansion valid for $r < r'$ - we obtain
\begin{equation}
\phi({\bf r})= \sum_{l=0}^{\infty} \frac{r^l}{a^{l+1}} \int_{r'=a}
d S' P_l({\bf e}_r'\cdot{\bf e}_r) f(\theta',\varphi')
\label{ecu23ac3}
\end{equation}
for $r < a$.  Using (\ref{ecu23ac2}) and (\ref{ecu23ac3}), the
jump condition for the normal derivative of the potential at any
point of the surface ${\bf r}=a {\bf e}_r(\theta,\varphi)\,$,
$\,4\pi f(\theta,\varphi)=\left[-\partial \phi/\partial
r\right]_{a^-}^{a^+}$, writes
\begin{equation}
f(\theta,\varphi)= \sum_{l=0}^{\infty} \frac{2l+1}{4\pi} \int
d\Omega' P_l({\bf e}_r'\cdot{\bf e}_r) f(\theta',\varphi'),
\label{ecu23ac4}
\end{equation}
where we have set $dS'=a^2 d\Omega'$.

Observe that if we replace each $P_l({\bf e}_r'\cdot{\bf e}_r)$ in
the RHS of (\ref{ecu23ac4}) by its expression as the second term
of (\ref{ecu23ad}) and carry out the integrals in the primed
quantities, we obtain an expansion of $f(\theta,\varphi)$ in terms
of (Maxwell's) surface harmonics. Therefore, any finite,
physically admissible, function defined on the unit sphere admits
an expansion in surface harmonics.  The completeness of any system
of orthonormal surface harmonics of the form $\left[ X_{ln} ({\bf
e}_r) : l=0,1,2...\, {\rm and}\, n=1,2...2l+1\right]$ now follows
by combining the results (\ref{ecu23ac4}) and (\ref{ecu23p}) as
\begin{equation}
f(\theta,\varphi)= \sum_{l=0}^{\infty} \sum_{n=1}^{2l+1} \int
d\Omega' f(\theta',\varphi') X^*_{ln}({\bf e}_r') X_{ln}({\bf
e}_r). \label{ecu23p1}
\end{equation}
Since $d\Omega'=\sin \theta' d\varphi' d\theta'$, equality of both
sides in (\ref{ecu23p1}) demands the so-called {\it completeness
relation}\cite{Luca}:
\begin{equation}
\frac{\delta(\varphi-\varphi')\delta(\theta-\theta')}{\sin
\theta}= \sum_{l=0}^{\infty} \sum_{n=1}^{2l+1} X^*_{ln}({\bf
e}_r') X_{ln}({\bf e}_r). \label{ecu23p2}
\end{equation}

If in (\ref{ecu23p1}) and (\ref{ecu23p2}) we replace the generic
$X_{ln}$'s by the standard $Y_{lm}$'s we obtain the familiar
results\cite{Jack,Sch}
\begin{equation}
f(\theta,\varphi)= \sum_{l=0}^{\infty} \sum_{m=-l}^{l} A_{lm}
Y_{lm}({\bf e}_r)\, \rightarrow \, A_{lm}=\int d\Omega'
f(\theta',\varphi') Y^*_{lm}({\bf e}_r') \label{ecu23p3}
\end{equation}
and
\begin{equation}
\frac{\delta(\varphi-\varphi')\delta(\theta-\theta')}{\sin
\theta}= \sum_{l=0}^{\infty} \sum_{m=-l}^{l} Y^*_{lm}({\bf e}_r')
Y_{lm}({\bf e}_r). \label{ecu23p4}
\end{equation}
Note that if $f$ is independent of $\varphi$ only the $A_{lm}$'s
with $m=0$ survive in (\ref{ecu23p3}) (since $Y_{lm}\sim
e^{im\varphi})$, and therefore in this particular case we can
write the expansion in terms of Legendre's polynomials. In effect,
using (\ref{ecu45}) and replacing ${\bf e}'_r$ by ${\bf e}_z$ in
(\ref{ecu23ad}), we find that the functions $Y_{l0}$
$P_l(\cos\theta)$ are proportional:
\begin{equation}
Y_{l0}({\bf r})= \sqrt{\frac{2l+1}{4\pi}} \frac{\,(-1)^{l} r^{l+1}
} {l!} \frac{\partial ^{l} (1/r)}{\partial
z^{l}}=\sqrt{\frac{2l+1}{4\pi}} P_l(\cos \theta), \label{ecu23ad1}
\end{equation}
equation (\ref{ecu23p3})  can be written as
\begin{equation}
f(\theta)= \sum_{l=0}^{\infty} B_l P_{l}(\cos \theta)\,
\rightarrow \, B_{l}=\frac{2l+1}{2} \int_{0}^{\pi} d\theta'
\sin\theta' f(\theta') P_{l}(\cos \theta'), \label{ecu23p5}
\end{equation}
where we have used the result
\begin{equation}
\int_{0}^{\pi} d\theta' \sin\theta'  P_{l}^2 (\cos
\theta')=\frac{1}{2\pi} \frac{4\pi} {2l+1} \int d\Omega'
Y^2_{l0}({\bf e}_r')=\frac{2} {2l+1}. \label{ecu23p6}
\end{equation}

As a first illustration of surface harmonics expansions, let us
consider the electrostatic field due a point charge $q$ located at
a point ${\bf R}$ external to an arbitrarily thin conducting
sphere of radius $a$ and grounded at zero potential. This problem
illustrates well the relation between the method of
images\cite{Jack} and that of surface harmonics
expansions\cite{Sch}. The potential at any point ${\bf r}$
external to the sphere is
\begin{equation}
\phi({\bf r})=\frac{q}{|{\bf r}-{\bf R}|}+\int_{r'=a} dS'
\frac{\sigma({\bf r}')}{|{\bf r}-{\bf r}'|}, \label{ecu23p6a}
\end{equation}
where $\sigma({\bf r}')$ is the {\it a priori} unknown surface
charge distribution on the sphere. To determine it, we substitute
the expansions $\sigma=\sum_{lm} \sigma_{lm} Y_{lm} ({\bf e'}_r)$,
$|{\bf r}-{\bf r}'|^{-1}= 4\pi/(2l+1) \sum_{lm} r'^l/r^{l+1}
Y_{lm} ({\bf e}_r) Y_{lm} ({\bf e'}_r)$ and $|{\bf r}-{\bf
R}|^{-1}= 4\pi/(2l+1) \sum_{lm} r^l/R^{l+1} Y_{lm} ({\bf e}_r)
Y_{lm} ({\bf e}_R)$ into (\ref{ecu23p6a}) and carry out the
surface integral using the orthogonality properties of the
spherical harmonics, which yields:
\begin{equation}
\phi({\bf r})=\frac{4\pi q}{2l+1} \sum_{lm} \frac{r^l}{R^{l+1}}
Y_{lm} ({\bf e}_r) Y_{lm} ({\bf e}_R)+\frac{4 \pi }{2l+1}\sum_{lm}
\frac{a^{l+2}}{r^{l+1}} \sigma_{lm} Y_{lm} ({\bf e}_r).
\label{ecu23p6b}
\end{equation}
Since $\phi=0$ at the sphere's surface, particularizing
(\ref{ecu23p6b}) for $r=a$ we obtain
\begin{equation}
\sigma_{lm}=-\frac{q}{a^2}\left(\frac{a}{R}\right)^{l+1}
Y_{lm}({\bf e}_R), \label{ecu23p6c}
\end{equation}
and (\ref{ecu23p6b}) can be written as:
\begin{equation}
\phi({\bf r})=\frac{q}{|{\bf r}-{\bf R}|}- \frac{4 \pi (a/R) q
}{2l+1}\sum_{lm} \frac{(a^2/R)^{l}}{r^{l+1}} Y_{lm} ({\bf e}_r)
Y_{lm} ({\bf e}_R)=\frac{q}{|{\bf r}-{\bf R}|}-\frac{(a/R)
q}{|{\bf r}-{\bf e}_R a^2/R |}. \label{ecu23p6d}
\end{equation}
Equation (\ref{ecu23p6d}) expresses the solution to the original
problem for potential at any point external to the sphere as the
superposition of those due to $q$ and an {\it image} (fictitious)
charge of magnitude $(a/R)q$ located inside the sphere at the
point ${\bf e}_R a^2/R$. Of course, if instead of a point charge
we have a charge density distribution $\rho({\bf R})$ within a
region $V$ external to the sphere, we must replace $q$ by
$\rho({\bf R}) d^3 R$ in (\ref{ecu23p6d}) and integrate over $V$.

As a second illustration, let us compute the electrostatic
interaction energy of two charge density distributions $\rho_1$
and $\rho_2$ occupying non-overlapping regions $V_1$ and $V_2$,
respectively.  Taking two points $O_1$ and $O_2$ within $V_1$ and
$V_2$, respectively, and denoting by ${\bf r}_1$ (${\bf r}_2$)
the position vectors with respect to $O_1$ ($O_2$) of a generic
point of $V_1$ ($V_2$), the interaction energy is
\begin{equation}
E=\int_{V_1} \int_{V_2} dV_1 dV_2 \frac{\rho_1({\bf r}_1)
\rho_2({\bf r}_2)}{|{\bf r}+{\bf r}_2-{\bf r}_1|}, \label{ecu23p7}
\end{equation}
where ${\bf r}$ is the position vector of $O_2$ with respect to
$O_1$. Since $|{\bf r}_1|$ and $|{\bf r}_2|$ are less than $|{\bf
r}|$ we can expand  $|{\bf r}+{\bf r}_2-{\bf r}_1|^{-1}$
independently in  ${\bf r}_1$ and in ${\bf r}_2$ by means of the
double Taylor series:
\begin{equation}
\frac{1}{|{\bf r}+{\bf r}_2-{\bf r}_1|}=\sum_{l_1 l_2} (-1)^{l_1}
r_2^{l_2} r_1^{l_1} \frac{({\bf e}_2\cdot\nabla)^{l_2}}{l_2!}
\frac{({\bf e}_1\cdot\nabla)^{l_1}}{l_1!} \frac{1}{r}.
\label{ecu23p8}
\end{equation}
But
\begin{equation}
 \frac{(-1)^{l_1}({\bf
e}_1\cdot\nabla)^{l_1}(1/r)}{l_1!}=\frac{P_{l_1}({\bf
e}_1\cdot{\bf e}_r)}{r^{l_1+1}}=\frac{4\pi}{2l_1+1}
\sum_{m_1=-l_1}^{l_1} C_{l_1m_1} Y^*_{l_1m_1}( {\bf e_1})
\frac{\partial ^{l_1} (1/r)}{\partial z^{l_1-m_1}
\partial \eta^{m_1}},
 \label{ecu23p9}
\end{equation}
where we have used the addition theorem (\ref{ecu23p}) and
expressed $Y_{l_1m_1}({\bf e}_r)$ in terms of $V_{l_1m_1}$ using
(\ref{ecu38}) and (\ref{ecu31}). Analogously,
\begin{equation}
\frac{(-1)^{l_2}({\bf
e}_2\cdot\nabla)^{l_2}(1/r)}{l_2!}=\frac{4\pi}{2l_2+1}
\sum_{m_2=-l_2}^{l_2} C_{l_2m_2} Y^*_{l_2m_2}( {\bf e_2})
\frac{\partial ^{l_2} (1/r)}{\partial z^{l_2-m_2}
\partial \eta^{m_2}}, \label{ecu23p10}
\end{equation}
whence applying the operator $({\bf e}_2\cdot\nabla)^{l_2}/l_2!$
to (\ref{ecu23p9})and commuting it with the derivatives in the
last term we obtain, on using (\ref{ecu23p10}),
\begin{equation}
(-1)^{l_1} \frac{({\bf e}_2\cdot\nabla)^{l_2}}{l_2!} \frac{({\bf
e}_1\cdot\nabla)^{l_1}}{l_1!} \frac{1}{r}=\frac{4^2\pi^2
(-1)^{l_2}}{(2l_2+1)(2l_1+1)} \sum_{m_2m_1} C_{l_1m_1} C_{l_2m_2}
Y^*_{l_1m_1}( {\bf e_1})Y^*_{l_2m_2}( {\bf e_2}) \frac{\partial
^{L} (1/r)}{\partial z^{L-M}
\partial \eta^{M}}, \label{ecu23p11}
\end{equation}
where $L \equiv l_1+l_2$ and $M \equiv m_1+m_2$. If we now
substitute (\ref{ecu23p11}) into (\ref{ecu23p8}) taking into
account that $Y_{LM}=C_{LM} r^{L+1} V_{LM}$ and defining
$A_{LM}\equiv \left[4^2\pi^2 (-1)^{l_2}C_{l_1m_1}
C_{l_2m_2}\right]/\left[(2l_2+1)(2l_1+1) C_{LM}\right]$, or
\begin{equation}
A_{LM}=\frac{4^2\pi^2
(-1)^{l_2}}{(2l_2+1)(2l_1+1)}\left[\frac{(2l_2+1)(2l_1+1)(L+M)!(L-M)!}{(2L+1)(l_1+m_1)!(l_1-m_1)!(l_2+m_2)!(l_2-m_2)!}\right]^{1/2},
\label{ecu23p12}
\end{equation}
we finally obtain
\begin{equation}
E= \sum_{l_1l_2m_1m_2} \frac{A_{LM}}{r^{L+1}}
(\rho_1)_{l_1m_1}Y_{LM}( {\bf e_r})(\rho_2)_{l_2m_2} ,
\label{ecu23p13}
\end{equation}
where we have defined the multipole moment, $\rho_{lm}$,  of a
charge density distribution occupying a region $V$ as
\begin{equation}
\rho_{lm}=\int_V dV' r'^l \rho({\bf r}') Y_{lm}( \bf {r}').
\label{ecu23p14}
\end{equation}
Except for a factor of $\sqrt{4\pi}$, expression (\ref{ecu23p13})
was obtained by Schwinger at al.\cite{Sch} using a different
reasoning.

\section{Partial wave expansions and
 Hobson's integral theorem}
Partial wave expansions are of fundamental importance in
scattering problems in both classical\cite{Sch} and quantum
physics\cite{Heit}. In this section we consider the expansions in
any orthonormal basis of surface or solid harmonics of the
function $e^{{\bf q}\cdot{\bf r}}$ - which leads to the partial
wave expansion of a plane wave, or Rayleigh's expansion\cite{Ray}
-, and for the translation operator\cite{Wenig2,Rowe} $e^{{\bf
q}\cdot \nabla}$, where ${\bf q}$ is a constant vector. We will
see that the latter expansion immediately leads to a proof of
Hobson's integral theorem\cite{Hob} mentioned in the introduction.
 Let us first consider the case in which ${\bf q}$ is a vector of the form ${\bf a}=q {\bf
e}_q$, where ${\bf e}_q$ is a real unit vector and the scalar $q$
may be complex. Then ${\bf q}\cdot{\bf r}=qr\cos\theta$, where
$\cos\theta\equiv{\bf e}_q\cdot{\bf e}_r$, and we can use
(\ref{ecu23p5}) to expand $e^{{\bf a}\cdot{\bf r}}$ in terms of
Legendre polynomials in the form


\begin{equation}
e^{{\bf q}\cdot{\bf r}}=\sum_{l=0}^{\infty} f_l(qr)
P_l(\cos\theta).\label{116}
\end{equation}
In order to determine the functions $f_l$ it is first convenient
to rewrite (\ref{116}) using Legendre's addition theorem
(\ref{ecu23p}) as
\begin{equation}
e^{{\bf q}\cdot{\bf r}}=\sum_{l=0}^{\infty} \frac{4\pi f_l(qr)
}{2l+1}\sum_{n=1}^{2l+1} X^*_{ln}({\bf e}_q) X_{ln}({\bf e}_r),
\label{117}
\end{equation}
where the $X_{ln}$'s are the members of any given orthonormal
basis of surface harmonics of degree $l$. In terms of the regular
solid harmonics $H_{ln}=r^l X_{ln}$ (\ref{117}) writes
\begin{equation}
e^{{\bf q}\cdot{\bf r}}=\sum_{l=0}^{\infty} \frac{4\pi f_l(qr)
}{(2l+1)q^lr^l}\sum_{n=1}^{2l+1} H^*_{ln}({\bf q}) H_{ln}({\bf r})
\label{117a}
\end{equation}

If we multiply(\ref{117a}) by $H_{ln}({\bf q})d\Omega_q$ and
integrate over the unit sphere using (\ref{120aa}) and taking into
account that $\int d\Omega_q H_{ln} H^{*}_{l'n'}=q^{l}q^{l'}\int
d\Omega X_{ln} X^{*}_{l'n'} =q^{2l}\delta_{ll'}\delta_{nn'}$  we
obtain
\begin{equation}
\frac{4\pi f_l(qr)}{2l+1} H_{ln}({\bf r})=4\pi H_{ln}({\bf r}) r^l
q^l s_l(qr), \label{118}
\end{equation}
where the function $s_l$ is defined in (\ref{122b}). Hence
(\ref{117}) takes the form
\begin{equation}
e^{{\bf q}\cdot{\bf r}}=4\pi\sum_{l=0}^{\infty} q^l r^l s_l(qr)
\sum_{n=1}^{2l+1} X^*_{ln}({\bf e}_q) X_{ln}({\bf e}_r).
\label{122a}
\end{equation}
In particular, for a plane wave one has  ${\bf q}=i {\bf k}$,
where ${\bf k}$ is a real vector, $qr= i k r$ and, since
$\sinh(ikr)=i\sin(kr)$, it is {\it customary} to rewrite
(\ref{122a}) in terms of {\it the spherical Bessel
function}\cite{Wenig2} of order $l$ defined as $j_l(\zeta)\equiv
\zeta^l s_l(i\zeta)$ or, using (\ref{122b}),
\begin{equation}
 j_l(\zeta) \equiv (-1)^{l} \zeta^l
\left(\frac{1}{\zeta}\frac{d}{d \zeta}\right)^l \left(\frac{\sin
\zeta}{\zeta}\right)   \label{122b1}
\end{equation}
 Then the expansion (\ref{122a}) yields the well known Rayleigh's\cite{Ray} expansion
for a plane wave:
\begin{equation}
e^{i {\bf k}\cdot{\bf r}}= \sum_{l=0}^{\infty} (2l+1) i^l j_l(kr)
P_l({\bf e}_k\cdot{\bf e}_r), \label{123}
\end{equation}
where we have replaced the sum over $n$ in  (\ref{122a}) using
Legendre's theorem (\ref{ecu23p}).

On the other hand, (\ref{122a}) writes in terms of the solid
harmonics $H_{ln}$ as

\begin{equation}
e^{{\bf q}\cdot{\bf r}}=4\pi\sum_{l=0}^{\infty}  s_l(qr)
\sum_{n=1}^{2l+1} H^*_{ln}({\bf q}) H_{ln}({\bf r}), \label{123aa}
\end{equation}
which is a good starting point to obtain the expansion for the
translation operator\cite{Wenig2} $e^{{\bf q}\cdot\nabla}$. In
effect, by formally replacing ${\bf r}$ in (\ref{123aa}) by the
operator $\nabla$ and $q r=q \sqrt{{\bf r}\cdot{\bf r}}$ by $q
\sqrt{\nabla^2}$ (see Section II) we obtain
\begin{equation}
e^{{\bf q}\cdot\nabla}=4\pi\sum_{l=0}^{\infty} \sum_{n=1}^{2l+1}
H^*_{ln}({\bf q}) H_{ln}({\bf \nabla})
s_l(q\sqrt{\nabla^2}),\label{122c}
\end{equation}
where we have used the fact that the operators $H_{ln}({\bf
\nabla})$ and  $s_l(q\sqrt{\nabla^2})$ obviously commute.

As a first illustration of (\ref{122c}), let us consider the
expansion of a function of the form $\phi (|{\bf r}-{\bf r}'|)$.
If, for concreteness, we assume $r > r'$,  Taylor's expansion
around $r=0$ can be written in the well known form $\phi (|{\bf
r}-{\bf r}'|)=e^{-{\bf r}' \cdot \nabla}\, \phi(r)$, whence
\begin{equation} \phi (|{\bf r}-{\bf r}'|)=
4\pi\sum_{l=0}^{\infty} \sum_{n=1}^{2l+1} (-1)^l H^*_{ln}({\bf
r'}) \left[H_{ln}({\bf \nabla}) s_l (r'\sqrt{\nabla^2})\right]
\phi(r),\label{122caa}
\end{equation}
where we have taken into account that $H^*_{ln}({\bf r'})=(-1)^l
H^*_{ln}({\bf r'})$
 An important particular case of (\ref{122caa}) is that of a
spherical wave \cite{ Jack,Sch} due to a point source located at
${\bf r'}$, $e^{ik|{\bf r}-{\bf r'}|}/|{\bf r}-{\bf r'}|$, where
$k$ is the wave number. In this case
\begin{equation}
\nabla^2\left(\frac{e^{ikr}}{r}\right)=-\frac{k^2e^{ikr}}{r},\label{130b}
\end{equation}
whence $s_l (r'\sqrt{\nabla^2}) \left(e^{ikr}/r\right)=s_l
(r'\sqrt{ -
k^2})\left(e^{ikr}/r\right)=s_l(ikr')\left(e^{ikr}/r\right)$. Thus
(\ref{122caa}) can be written as
\begin{equation}
\frac{e^{ik|{\bf r}-{\bf r'}|}}{|{\bf r}-{\bf r'}|}=
4\pi\sum_{l=0}^{\infty} \sum_{n=1}^{2l+1} (-1)^l s_l(ikr')
H^*_{ln}({\bf r'}) H_{ln}({\bf \nabla})\left(\frac{e^{ikr}}{r}
\right) \label{130b1}
\end{equation}
or, using (\ref{ecu9ad4}),
\begin{equation}
\frac{e^{ik|{\bf r}-{\bf r'}|}}{|{\bf r}-{\bf r'}|}=
4\pi\sum_{l=0}^{\infty} \sum_{n=1}^{2l+1} (-1)^l
s_l(ikr')H^*_{ln}({\bf r'}) H_{ln}({\bf r})
\left(\frac{1}{r}\frac{d}{d
r}\right)^{l}\left(\frac{e^{ikr}}{r}\right) . \label{130b1}
\end{equation}
It is {\it customary} to write (\ref{130b1}) in terms of the
spherical Hankel function of the first kind and of order $l$
defined as
\begin{equation}
h^{(1)}_l(\zeta)\equiv\frac{(-1)^l\zeta^l}{i}\left(\frac{1}{\zeta}\frac{d}{d\zeta}\right)^l
\left(\frac{e^{i\zeta}}{\zeta}\right), \label{130b2}
\end{equation}
in which case, using the definition $j_l(kr')=k^l r'^l s_l(ikr')$,
substituting $H^*_{ln}({\bf r'}) H_{ln}({\bf r})=r'^l r^l
X^*_{ln}({\bf e}_r') X_{ln}({\bf e}_r)$ and taking account that
for $r < r'$ an analysis entirely analogous to the one carried out
above leads  to an expression of the same form as (\ref{130b1})
but with the roles of ${\bf r}$ and ${\bf r'}$ interchanged,  the
result (\ref{130b1}) can be expressed in the more familiar form
\begin{equation}
\frac{e^{ik|{\bf r}-{\bf r'}|}}{|{\bf r}-{\bf r'}|} =4\pi i k
\sum_{l=0}^{\infty} \sum_{j=0}^{n=2l+1} j_l(kr_<) h^{(1)}_l(kr_>)
X^*_{ln}({\bf r}_<) X_{ln}({\bf r}_>). \label{130b4}
\end{equation}
The result (\ref{130b4}) constitutes the general expression for
the expansion of a spherical wave in any basis of orthonormal
surface harmonics.

As a second application of  (\ref{122c}) we will provide an
apparently new proof of Hobson's theorem\cite{Hob} for the
integral over a sphere of radius $R$ of the product of a surface
harmonic, $Y_k({\bf r})$, and any function $F({\bf r})$ which is
analytic for $r\le R$. In effect, by replacing in (\ref{122c})
vector ${\bf a}$ by ${\bf r}$ and $a^2$ by $R^2$, Taylor's
expansion around the origin for $F({\bf r})$ can be written as,
\begin{equation} F({\bf r})=e^{{\bf r} \cdot \nabla_o} F=
4\pi\sum_{l=0}^{\infty} \sum_{n=1}^{2l+1}  H^*_{ln}({\bf r})
\left[H_{ln}({\bf \nabla}) s_l(R\sqrt{\nabla^2})\right]_o
F,\label{122ca}
\end{equation}
where subscript $o$ means that the derivatives of $F$ must be
particularized at the origin. Now, if we normalize the surface
harmonic $Y_k$ by dividing it by $\sqrt{C}$ - where $C\equiv \int
d\Omega Y_k Y^*_k$ -, then $Y_k/\sqrt{C}$ can always be taken as
one of the elements of the generic basis $X_{ln}$, in which case
$r^k Y_k/\sqrt{C}$ replaces one of the $H_{ln}$ in (\ref{122ca}).
The rest of the basis elements could be obtained, for example, via
Gram-Schmidt orthogonalization. Therefore, since $Y_k$ is
orthogonal to all solid harmonics in (\ref{122ca}) except to
$H_k/\sqrt{C}$ and $\int dS Y_k H_k^* = R^{k+2}C$, we obtain
\begin{equation} \int_{r=R} dS Y_k F({\bf r})=
4\pi R^{k+2} \left[s_k(R \sqrt{\nabla^2}) H_k({\bf
\nabla})\right]_o F,\label{122cb}
\end{equation}
which is Hobson's integral theorem.

\section{The rotation matrix for the spherical harmonics.}

It is well known\cite{Bied,Wig} that if a given rotation brings
the position vector ${\bf r}$ into the vector  ${\bf r}'$ - the
{\it image} of ${\bf r}$-, then the value at ${\bf r}'$ of any
spherical harmonic of degree $l$, $Y_{lm'}({\bf r}')$, is a linear
combination of the values at ${\bf r}$ of all $2l+1$ spherical
harmonics of the set of degree $l$, i.e. $Y_{lm'}({\bf r'})=
\sum_{m=-l}^{l} D^l_{m'm} Y_{lm}({\bf r})\quad (m'=-l,...l)$,
where the elements of the {\it rotation matrix}, $D^l_{m'm}$, are
uniquely determined  by the rotation. In this section we will
carry out an analysis of the rotational properties of the
spherical harmonics based on the invariance under rotations of the
scalar product, that is, if ${\bf r}'$ and ${\bf b}'$ are the
images of vectors ${\bf r}$ and ${\bf b}$, then $\left({\bf
b}\cdot {\bf r}\right)^l=\left({\bf b}'\cdot {\bf r}'\right)^l$
or, since $r=r'$, $\left({\bf b}\cdot {\bf
e}_r\right)^l=\left({\bf b}'\cdot {\bf e}'_r\right)^l$. Thus if
${\bf b}$ is a null vector, so is its image ${\bf b}'\,$ (${\bf
b}^2={\bf b}'^2=0$) and $\left({\bf b}\cdot {\bf r}\right)^l$ and
$\left({\bf b}'\cdot {\bf r}\right)^l$ are solid harmonics whose
values at ${\bf e}_r$ and ${\bf e}'_r$ can be expanded in terms of
the $Y_{lm}({\bf e}_r)$ and $Y_{lm}({\bf e}'_r)$ ($\quad
(m=-l,...l)$, respectively. Then the elements of the rotation
matrix are obtained after equating corresponding terms in both
expansions.

According to (\ref{ecu23p3}) we can write
\begin{equation}
\left({\bf b}\cdot {\bf e}_r\right)^l = \sum_{l=0}^{\infty}
\sum_{m=-l}^{l} Y_{lm}({\bf e}_r) \int d\Omega'' \left({\bf
b}\cdot {\bf e}''_r\right)^l Y^*_{lm}({\bf e}''_r),
\label{ecu26am1}
\end{equation}
where the double prime denotes the generic point on the unit
sphere. Since, on the unit sphere $\left({\bf b}\cdot {\bf
r}\right)^l=\left({\bf b}\cdot {\bf e}_r\right)^l$, the integrals
on the RHS of (\ref{ecu26am1}) can be computed using Maxwell's
theorem (\ref{ecu13adc}) with
\begin{equation}
a\equiv 1,\quad \Phi\equiv\left({\bf b}\cdot{\bf
r}\right)^l\quad{\rm and}\quad Y_l\equiv
Y^*_{lm}=\sqrt{\frac{2l+1}{4\pi}} \frac{\,(-1)^{l+m}
2^{m}\,r^{l+1} } {\sqrt{(l+m)!(l-m)!}} \frac{\partial ^{l}
(1/r)}{\partial z^{l-m} \partial \xi^m} \label{ecu26a}
\end{equation}
which yields
\begin{equation}
\int d\Omega''  \left({\bf b}\cdot {\bf e}''_r\right)^l
Y^*_{lm}({\bf e}''_r)=\sqrt{\frac{4\pi}{2l+1}}\,\frac{\,(-1)^{m}
2^{m}}{\sqrt{(l+m)!(l-m)!}} \frac{\partial ^l\left({\bf
b}\cdot{\bf r}\right)^l}{\partial z^{l-m}\partial
\xi^m}\Big|_{r=0}. \label{ecu98}
\end{equation}

In order to compute the RHS of (\ref{ecu98}), we express ${\bf b}$
in the spherical basis,
\begin{equation}
{\bf b}=b_\xi {\bf e}_\xi + b_\eta {\bf e}_\eta + b_z {\bf e}_z ,
\label{ecu72aa3}
\end{equation}
and dot it with ${\bf r}=(\xi {\bf e}_\xi + \eta {\bf
e}_\eta)/\sqrt{2} + z {\bf e}_z$, which yields
\begin{equation}
({\bf b}\cdot{\bf r})^l=\left[(b_\xi\eta+b_\eta\xi)/\sqrt{2}+z
b_z\right]^l=\sum_{0}^{l} \frac{l!2^{-j/2}}{(l-j)!j!}
(b_\xi\eta+b_\eta\xi)^{j} z^{l-j} b_z^{l-j}. \label{ecu72aa5}
\end{equation}
Thus, since the only term in the binomial expansion of $({\bf
b}\cdot{\bf r})^l$ which contributes to the derivative in the RHS
of  (\ref{ecu98}) is that containing the product $z^{l-m} \xi^m$-
whose coefficient is readily found from (\ref{ecu72aa5}) to be
$l!2^{-m/2}\, b_\eta^{m} b_z^{l-m}/[(l-m)!m!]$- we have
\begin{equation}
\frac{\partial ^l\left({\bf b}\cdot{\bf r}\right)^l}{\partial
z^{l-m}\partial \xi^m}\Big|_{r=0}=l!\,2^{-m/2}\, b^m_\eta
b_z^{l-m}=l!\,2^{l/2-m}\, i^{l-m} b_\xi^{(l-m)/2}
b_\eta^{(l+m)/2},
 \label{ecu98a}
\end{equation}
where we have used the fact that
\begin{equation}
{\bf b}^2=0\rightarrow b_z=i\sqrt{\,2 b_\xi b_\eta}\quad {\rm or}
\quad {\bf b}\cdot {\bf e}_z=i\sqrt{\,2 {\bf b}\cdot {\bf e}_\eta
\,{\bf b}\cdot {\bf e}_\xi}. \label{ecu72aa4}
\end{equation}
 Substituting
(\ref{ecu98a}) into (\ref{ecu98}) we finally obtain the expansion
\begin{equation}
\left({\bf b}\cdot{\bf r}/r\right)^l=2^{l/2} i^l
\,l!\sqrt{\frac{4\pi }{2l+1}}\,\sum_{m=-l}^{m=l} \frac{i^m
b_\xi^{(l-m)/2}b_\eta^{(l+m)/2} Y_{lm}({\bf
e}_r)}{\sqrt{(l+m)!(l-m)!}},   \label{ecu52f}
\end{equation}
which was introduced  by Kramers\cite{Kra} and, later, by
Schwinger\cite{Sch} as a generating function for {\it defining}
the spherical harmonics on the LHS. Here, we have preferred to
proceed instead by expanding $\left({\bf b}\cdot{\bf
r}/r\right)^l$ into the already known [see (\ref{ecu45})]
$Y_{lm}$'s and, then, finding the expansion coefficients using
Maxwell's integral theorem.

 The invariance under rotations of (\ref{ecu52f}) makes it a
suitable expression to analyze the rotational transformation
properties of the spherical harmonics. In effect, let there be
given two orthonormal, righthanded sets of vectors, $\left({\bf
e}_x, {\bf e}_y, {\bf e}_z={\bf e}_x\times {\bf e}_y\right)$ and
$\left({\bf e}'_x, {\bf e}'_y, {\bf e}'_z={\bf e}'_x\times {\bf
e}'_y\right)$, and consider the rotation about the origin of
coordinates defined as that bringing
 the vector ${\bf e}_x$ into ${\bf e}'_x$ {\it and} the vector
${\bf e}_y$ into ${\bf e}'_y$. Then, any vector which rotates
rigidly with the unprimed system will be brought by the rotation
into another vector - the image - that has the same components
with respect to the primed system as has the former with respect
to the primed one. Hence if with respect to the spherical basis
associated with the unprimed set a vector ${\bf b}$ has components
$b_\xi$, $b_\eta$ and $b_z$ and its image, ${\bf b}'$, has
components $b'_\xi$, $b'_\eta$ and $b'_z$ , we have
\begin{equation} {\bf b}'= b'_\xi {\bf e}_\xi  +
b'_\eta {\bf e}_\eta + b'_z {\bf e}_z= b_\xi {\bf e}'_\xi  +
b_\eta {\bf e}'_\eta + b_z {\bf e}'_z , \label{ecu52g}
\end{equation}
where ${\bf e}'_\xi$, ${\bf e}'_\eta$ and ${\bf e}'_z$ are the
vectors of the spherical basis associated with the primed set. On
taking the scalar product of both sides of the last equality in
(\ref{ecu52g}) by ${\bf e}'_\eta$ and ${\bf e}'_\xi$,
respectively, we obtain
\begin{equation} b_\xi = b'_\xi {\bf
e}'_\eta\cdot {\bf e}_\xi  + b'_\eta {\bf e}'_\eta \cdot{\bf
e}_\eta + b'_z {\bf e}'_\eta\cdot {\bf e}_z, \label{ecu52ga1}
\end{equation}
and
\begin{equation} b_\eta = b'_\xi {\bf
e}'_\xi\cdot {\bf e}_\xi  + b'_\eta {\bf e}'_\xi \cdot{\bf e}_\eta
+ b'_z {\bf e}'_\xi\cdot {\bf e}_z . \label{ecu52ga2}
\end{equation}
Now, if ${\bf b}$ is a null vector, so is its image under
rotation, ${\bf b}'$, whence
\begin{equation}
b'_z=i\sqrt{\,2 b'_\xi b'_\eta}\label{ecu52gb1} \end{equation}
and, since ${\bf e}'_\eta$ and ${\bf e}'_\xi$ are null vectors,
(\ref{ecu72aa4}) yields
\begin{equation}
{\bf e}'_\eta\cdot {\bf e}_z=i\sqrt{\,2 {\bf e}'_\eta\cdot {\bf
e}_\eta {\bf e}'_\eta \cdot {\bf e}_\xi}\,\quad {\rm and} \quad
{\bf e}'_\xi\cdot {\bf e}_z=-i\sqrt{\,2 {\bf e}'_\xi\cdot {\bf
e}_\eta \,{\bf e}'_\xi \cdot {\bf e}_\xi},  \label{ecu52gb2}
\end{equation}
where we have taken different branches for the square root in
order to satisfy the condition ${\bf e}'_\xi\cdot {\bf
e}_z=\left({\bf e}'_\eta \cdot {\bf e}_z\right)^*$. Introducing
(\ref{ecu52gb1}) and (\ref{ecu52gb2}) into (\ref{ecu52ga1}) and
(\ref{ecu52ga2}) it is easily seen that the transformation laws
for the $\xi$ and $\eta$ components of a null vector can be
written as the perfect squares
\begin{equation} b_\xi=\left(b'^{1/2}_\xi\sqrt{{\bf e'}_\eta\cdot{\bf
e}_\xi}-b'^{1/2}_\eta\sqrt{{\bf e'}_\eta\cdot{\bf
e}_\eta}\right)^2 \label{ecu52la}
\end{equation}
and
\begin{equation} b_\eta=\left(b'^{1/2}_\xi\sqrt{{\bf e'}_\xi\cdot{\bf
e}_\xi}+b'^{1/2}_\eta\sqrt{{\bf e'}_\xi\cdot{\bf
e}_\eta}\right)^2. \label{ecu52m}
\end{equation}
Thus, defining the quantities $s\equiv b^{1/2}_\xi$, $t\equiv
b^{1/2}_\eta$, $c\equiv\sqrt{{\bf e'}_\eta\cdot{\bf e}_\xi}$ and
$d\equiv\sqrt{{\bf e'}_\xi\cdot{\bf e}_\xi}$, we can write
(\ref{ecu52la})-(\ref{ecu52m}) more compactly as
\begin{equation} s = c s'- d^* t', \label{ecu52n}
\end{equation}
\begin{equation} t = d s' + c^* t'. \label{ecu52o}
\end{equation}
It is easy to show that (\ref{ecu52n})-(\ref{ecu52o}) is an
unitary transformation with unit determinant - i.e. it belongs to
the group SU2-, so that the quantities $b^{1/2}_\xi$ and
$b^{1/2}_\eta$ transform as the components of a {\it two-
spinor}\cite{Dowker,Kra}. In effect, from the definitions of $c$
and $d$, and according to (\ref{ecu52gb2}) we have
$cdc^*d^*=\sqrt{{\bf e'}_\xi\cdot{\bf e}_\eta{\bf e'}_\xi\cdot{\bf
e}_\xi {\bf e'}_\eta\cdot{\bf e}_\xi{\bf e'}_\eta\cdot{\bf
e}_\eta}={\bf e}'_\eta\cdot{\bf e}_z{\bf e}'_\xi\cdot{\bf e}_z/2$
and, therefore,
\begin{equation}
(cc^*+dd^*)^2={\bf e}'_\eta\cdot{\bf e'}_\xi{\bf e'}_\xi\cdot{\bf
e}_\eta+{\bf e'}_\xi\cdot{\bf e}_\xi{\bf e'}_\eta\cdot{\bf
e}_\eta+{\bf e}'_\eta\cdot{\bf e}_z{\bf e}'_\xi\cdot{\bf e}_z={\bf
e}'_\eta\cdot{\bf I}\cdot{\bf e}'_\xi=1, \label{ecu52p}
\end{equation}
where
\begin{equation}
{\bf I}={\bf e}_x{\bf e}_x+{\bf e}_y{\bf e}_y+{\bf e}_z{\bf
e}_z={\bf e}_\xi{\bf e}_\eta+{\bf e}_\eta{\bf e}_\xi+{\bf e}_z{\bf
e}_z \label{ecu52p}\end{equation} is the unit dyadic. Since we
assume that no reflections are involved in the transformation of
the coordinate system we take the positive root in (\ref{ecu52p})
and obtain $cc^*+dd^*=1$, whence the transformation
(\ref{ecu52n})-(\ref{ecu52o}) is unimodular.

Now, if the images under the rotation of the null vector ${\bf b}$
and of the position vector ${\bf r}$ are ${\bf b}'$ and ${\bf
r}'$, respectively, an expansion analogous to (\ref{ecu52f}) must
hold for $\left({\bf b}'\cdot{\bf r}'/r'\right)^l$ where $b_\xi$,
$b_\eta$ and ${\bf e}_r$ are replaced by $b'_\xi$, $b'_\eta$ and
${\bf e}'_r$. Thus, equating both expressions we obtain
\begin{equation}
\sum_{m'=-l}^{l} \frac{ i^{m'} s'^{l-m'} t'^{l+m'} Y_{lm'}({\bf
e}'_r)}{\sqrt{(l+m')!(l-m')!}} = \sum_{m=-l}^{l} \frac { i^m
s^{l-m} t^{l+m} Y_{lm}({\bf e}_r)}{\sqrt{(l+m)!(l-m)!}}.
\label{ecu52q}
\end{equation}
To find $Y_{lm'}({\bf e}'_r)$ in terms of the $Y_{lm}({\bf
e}_r)$'s we just have to substitute the transformation
(\ref{ecu52n})-(\ref{ecu52o}) into the RHS of (\ref{ecu52q}) and,
for each $m$, find the term containing $s'^{l-m'} t'^{l+m'}$. The
factor multiplying $s'^{l-m'} t'^{l+m'}$ in the $m$-th summand can
be computed as
\begin{equation}
\frac{1}{(l-m')!(l+m')!} \frac{\partial^{l-m'}}{\partial
s'^{l-m'}} \frac{\partial^{l+m'}}{\partial t'^{l+m'}}\left[(c s'-
d^* t')^{l-m}(d s' + c^* t')^{l+m}\right]_{(s',t')=(0,0)},
\label{ecu52r}
\end{equation}
or, applying Leibnitz rule to the $(l+m')$-th partial derivative
with repect to $t'$ of the product in (\ref{ecu52r}) and
evaluating the result at $t'=0$,
\begin{equation}
\frac{1}{(l-m')!} \frac{\partial^{l-m'}}{\partial s'^{l-m'}}
\sum_{k=0}^{l+m'}\frac{(l-m)!(l+m)!(-1)^k d^{*\,k}c^{*\,l+m'-k}
c^{l-m-k}d^{m-m'+k}s'^{l-m'}}{k!(l+m'-k)!(l-m-k)!(m-m'+k)!},
\label{ecu52s}
\end{equation}
with the convention that the factorial of a negative number is
infinity. Carrying out now the derivative with respect to $s'$ in
(\ref{ecu52s}) and substituting the result in (\ref{ecu52q}) we
readily obtain
\begin{equation}
Y_{lm'}(\theta',\varphi')= \sum_{m=-l}^{l} D^l_{m'm}({\bf e'}_\eta
, {\bf e}_\eta) Y_{lm}(\theta,\varphi),\label{ecu52t}
\end{equation}
where $(\theta',\varphi')$ and $(\theta,\varphi)$ are the
spherical coordinates of the points ${\bf e}'_r$ and ${\bf e}'_r$,
respectively, and we have set
\begin{equation}
\begin{split}
D^l_{m'm}({\bf e'}_\eta,  {\bf e}_\eta)\equiv
&i^{m-m'}\sqrt{\frac{(l-m)!(l+m)!}{(l-m')!(l+m')!}}\times
 \\ \sum_{k=0}^{l+m'}
\begin{pmatrix}l-m'\\m-m'+k\end{pmatrix}&\begin{pmatrix}l+m'\\k\end{pmatrix}
(-1)^k d^{*\,k}c^{*\,l+m'-k} c^{l-m-k}d^{m-m'+k} \label{ecu52u}
\end{split}
\end{equation}
with
\begin{equation}
c\equiv\sqrt{{\bf e'}_\eta\cdot{\bf e}_\xi}=\sqrt{{\bf
e'}_\eta\cdot{\bf e}^*_\eta},\quad {\it and}\quad
d\equiv\sqrt{{\bf e'}_\xi\cdot{\bf e}_\xi}=\sqrt{{\bf
e'}^*_\eta\cdot{\bf e}^*_\eta}.\label{ecu52v}
\end{equation}
Expression (\ref{ecu52u}) yields the elements of the rotation
matrix, $D^l_{m'm}({\bf e'}_\eta , {\bf e}_\eta)$, for the
rotation of the coordinate system defined as that bringing the
spherical vector ${\bf e}_\eta$ into ${\bf e}'_\eta$ - or,
equivalently, the vector ${\bf e}_x$ into ${\bf e}'_x$ {\it and}
the vector ${\bf e}_y$ into ${\bf e}'_y$-. Notice that
(\ref{ecu52u})-(\ref{ecu52v}) are independent of the particular
parametrization chosen to describe the rotation, and that
(\ref{ecu52v}) provides with a direct geometrical meaning for
parameters $c$ and $d$ of the SU2 transformation (\ref{ecu52n})-
(\ref{ecu52o}). In particular, it is easy to show that for the
usual Euler angles employed in Quantum Mechanics\cite{Bied} to
parametrize a rotation, namely: a) a rotation by an angle $\alpha$
around the z-axis, b) a rotation by an angle $\beta$ around the
new y-axis and c) a rotation by an angle $\gamma$ around the new
z-axis, one obtains (by expressing the vectors $[{\bf e}'_x, {\bf
e}'_y]$ in terms of $[{\bf e}_x, {\bf e}_y]$ and the Euler angles)
\begin{equation} ({\bf e'}_\eta\cdot{\bf
e}^*_\eta)^{1/2}=e^{-i(\alpha+\gamma)/2}\cos(\beta/2),\quad({\bf
e'}^*_\eta\cdot{\bf
e}^*_\eta)^{1/2}=e^{i(\gamma-\alpha)/2}\sin(\beta/2),
\label{ecu52w}
\end{equation}
and (\ref{ecu52u}) goes into the standard form of the rotation
matrix elements\cite{Wig}.

Finally, note that we have considered here only the case of
rotation matrices of integer rank. However, in the theory of
quantum angular momentum\cite{Bied} both integer and half-integer
ranks are, of course, of importance. A very complete presentation
of compact, parametrization-free representations of finite
rotation matrices of arbitrary rank using invariant tensor forms
and spinor operators can be found in the works by Manakov et
al.\cite{Mana1,Mana2,Mana3}, where they also give some
applications such as photo-emission by polarized atoms or
reduction formulae for some tensor constructions characterizing
photoprocesses.

\section{Gaunt integrals and Wigner coefficients.}

In this section we will apply Hobson's integral theorem
(\ref{122cb}) to obtain a generating function for integrals of
products of three spherical harmonics - the so-called Gaunt
integrals \cite{Gaunt}- and provide their expressions in terms of
the well known Wigner coefficients\cite{ Bied}. For this purpose,
let us first use (\ref{ecu52f}) to express the regular solid
harmonic associated to the spherical harmonic $Y_{lm}$,
$H_{lm}=r^lY_{lm}$,  as
\begin{equation} H_{lm}({\bf
r})=\sqrt{\frac{(2l+1)(l+m)!(l-m)!}{4\pi (-1)^{l+m} 2^{2l} l!^2}}
\left[\left({\bf b}\cdot{\bf r}\right)^l\right]_{s^{l-m}t^{l+m}},
\label{ecu52wa}
\end{equation}
where the subscript on the RHS means the coefficient of
$s^{l-m}t^{l+m}\equiv b_{\xi}^{(l-m)/2}b_{\eta}^{(l+m)/2}$ in the
binomial expansion of $\left({\bf b}\cdot{\bf r}\right)^l$. Thus
we have
\begin{equation}
\int_{r=R} dS
H_{l_bm_b}H_{l_cm_c}H_{l_dm_d}=\sqrt{\frac{(2l_b+1)(2l_c+1)(2l_d+1)\mu_b!\mu_c!\mu_d!\nu_b!\nu_c!\nu_d!}{4\pi
(-1)^{M} 2^{2L} l!_b^2 l!_c^2 l!_c^2}} I_{\rho_b\rho_c\rho_d},
\label{ecu52wb}
\end{equation}
where \begin{equation} I\equiv \int d\Omega \left({\bf b}\cdot{\bf
r}\right)^{l_b} \left({\bf c}\cdot{\bf r}\right)^{l_c} \left({\bf
d}\cdot{\bf r}\right)^{l_d}, \label{ecu52wb1}\end{equation} and,
to simplify notation, we have written
\begin{equation} \mu\equiv l-m,\quad \nu\equiv l+m,\quad
\rho\equiv {s^{l-m}t^{l+m}},\nonumber\end{equation}
\begin{equation} L\equiv l_b+l_c+l_d\quad{\rm and}\quad
M\equiv m_b+m_c+m_d. \label{ecu52wba}\end{equation}

In order to compute (\ref{ecu52wb1}) we apply Hobson's integral
theorem (\ref{122cb}) with $R=1$, $Y_k=H_k\equiv ({\bf b\cdot
r})^{l_b} $ and   $F({\bf r})\equiv ({\bf c\cdot r})^{l_c}({\bf
d\cdot r})^{l_d}$, whence
\begin{equation} I=
= 4\pi \left[s_{l_b}(\sqrt{\nabla^2}) ({\bf b\cdot \nabla})^{l_b}
\right]_o \left[({\bf c\cdot r})^{l_c}({\bf d\cdot
r})^{l_d}\right]. \label{122cbaa}
\end{equation}
 Since $({\bf c\cdot r})^{l_c}({\bf d\cdot r})^{l_d}$ is a
homogeneous polynomial of degree $l_c+l_d$, it is clear that the
RHS of (\ref{122cbaa}) vanishes unless $\lambda_b\equiv
l_c+l_d-l_b$ is zero or a positive even integer. Therefore, the
only term in the power series (\ref{122b}) for $S_{l_b}(\nabla^2)$
which contributes to (\ref{122cbaa})is that containing
$(\nabla^2)^{\lambda_b/2}$, which corresponds to $j= L/2$, where
$L\equiv l_c+l_d+l_b$. Thus, taking into account (\ref{122b}),
(\ref{122cbaa}) yields
\begin{equation}
I=\frac{4\pi
2^{l_b}(L/2)!}{(L+1)!(\lambda_b/2)!}\left[(\nabla^2)^{\lambda_b/2}
({\bf b \cdot \nabla})^{l_b} \right]_o \left[({\bf c\cdot
r})^{l_c}({\bf d\cdot r})^{l_d}\right].\label{122cbb}
\end{equation}
Now, since $\nabla^2({\bf c\cdot r})^{l_c}=\nabla^2({\bf d\cdot
r})^{l_d}=0$ we have
\begin{equation}
\nabla^2 \left[({\bf c\cdot r})^{l_c}({\bf d\cdot
r})^{l_d}\right]=2 \nabla({\bf c\cdot r})^{l_c}\cdot\nabla({\bf
d\cdot r})^{l_d}=2 l_c l_d\, {\bf c\cdot d}\,({\bf c\cdot
r})^{l_c-1}({\bf d\cdot r})^{l_d-1},\label{122cbc}
\end{equation}
and  repeating the process $\lambda_b/2$ times we easily obtain
\begin{equation}
(\nabla^2)^{\lambda_b/2}\left[({\bf c\cdot r})^{l_c}({\bf d\cdot
r})^{l_d}\right]= \frac{2^{\lambda_b/2}l_c!l_d!({\bf c\cdot
d})^{\lambda_b/2}}{(\lambda_c/2)!(\lambda_d/2)!} ({\bf c\cdot
r})^{\lambda_d/2} ({\bf d\cdot r})^{\lambda_c/2},\label{122cbdddd}
\end{equation}, where we have written $\lambda_c\equiv
l_d+l_b-l_c$ and $\lambda_d\equiv l_b+l_c-l_d$. Thus,
\begin{equation} I=
\frac{4\pi
2^{L/2}(L/2)!l_c!l_d!}{(L+1)!(\lambda_b/2)!(\lambda_c/2)!(\lambda_d/2)!}\left[
({\bf b \cdot \nabla})^{l_b} \right]_o \left[({\bf c\cdot
r})^{\lambda_d/2} ({\bf d\cdot
r})^{\lambda_c/2}\right].\label{122cbd}
\end{equation}
Also, Leibnitz's rule yields
\begin{equation}
({\bf b}\cdot \nabla)^{l_b}\left[({\bf c\cdot
r})^{\lambda_d/2}({\bf d\cdot
r})^{\lambda_c/2}\right]=\sum_{j=0}^{l_b} \frac{l_b!}{j!(l_b-j)!}
\left[({\bf b}\cdot \nabla)^{j}({\bf c\cdot
r})^{\lambda_d/2}\right]\,({\bf b}\cdot \nabla)^{l_b-j}({\bf
d\cdot r})^{\lambda_c/2},\label{122cbe}
\end{equation}
and the only term in the series in the LHS that gives non zero
contribution at ${\bf r=0}$ corresponds to $j=\lambda_d/2$, which
also makes $l_b-j=\lambda_c/2$. Thus
\begin{equation}
\left[({\bf b}\cdot \nabla)^{l_b}\right]_o \left[({\bf c\cdot
r})^{\lambda_d/2}({\bf d\cdot r})^{\lambda_c/2}\right]= l_b!({\bf
b\cdot c})^{\lambda_d/2}({\bf b\cdot
d})^{\lambda_c/2},\label{122cbf}
\end{equation}
and inserting (\ref{122cbf}) and (\ref{122cbd}) into
(\ref{122cbb})  we finally obtain
\begin{equation} \begin{split} I =
 \frac{4\pi 2^{L/2} (L/2)!l_b!l_c!l_d!({\bf c\cdot
d})^{\lambda_b/2} ({\bf b\cdot c})^{\lambda_d/2} ({\bf b\cdot
d})^{\lambda_c/2}}
{(L+1)!(\lambda_b/2)!(\lambda_c/2)!(\lambda_d/2)!}.\label{122cba}
\end{split}
\end{equation}
Notice that (\ref{122cba}) vanishes if $\lambda_b$ is odd or if
any of the parameters $\lambda_b$, $\lambda_c$ or $\lambda_d$.
Thus, given the values of $l_c$ and $l_d$ , $I\neq 0$ only if the
value $l_b$ is one of the set
\begin{equation}
l_c+l_d,\quad l_c+l_d-2,\quad ...\quad |lc-l_d|. \label{122cbda}
\end{equation}
The reader familiar we group theory might have noticed that this
set of integers is the same as that in the Clebsch-Gordan
series\cite{Weyl} for the decomposition into irreducible
representations of the tensor product of two irreducible
representations. Also, observe that the foregoing procedure based
on Hobson's integral theorem can be extended, in principle, to
compute integrals of products of any number of simple solid
harmonics $({\bf b\cdot r})^{l_b}({\bf c\cdot r})^{l_c}...({\bf
q\cdot r})^{l_q}$ but, of course, the result of applying the
operator $\left[(\nabla^2)^{(lc+..l_q-l_b)/2} ({\bf b\cdot
\nabla})^{l_b} \right]_o$ gets more involved as the number of
factors increases.

In order to compute $I_{\rho_b\rho_c\rho_d}$ we must find the
coefficient of the term
$s_b^{\mu_b}t_b^{\nu_b}s_c^{\mu_c}t_c^{\nu_c}s_d^{\mu_d}t_d^{\nu_d}$
in the expansion of the product $({\bf c\cdot
d})^{\lambda_b/2}({\bf b\cdot c})^{\lambda_d/2}({\bf b\cdot
d})^{\lambda_c/2}$ where $\lambda_b=l_c+l_d-l_b$,
$\lambda_c=l_d+l_b-l_c$ and $\lambda_b=l_b+l_c-l_d$. Note that if
we write ${\bf b}=b_\xi {\bf e}_\xi + b_\eta {\bf e}_\eta + b_z
{\bf e}_z$, $s_b\equiv b_\xi^{1/2}$, $t_b\equiv b_\eta^{1/2}$ and
$b_z\equiv i \sqrt{2} s_b t_b$ (${\bf b}$ null vector), with
corresponding definitions for $s_c$, $t_c$, $s_d$ and $t_d$, we
have
\begin{equation}
{\bf b\cdot c}=s_b^2 t_c^2 + t_b^2 s_c^2 - 2 s_b t_b s_c t_c =
(s_b t_c - t_b s_c)^2 \label{ecu52wc} \end{equation} and,
analogously,
\begin{equation}
{\bf b\cdot d}=(s_b t_d - t_b s_d)^2\quad{\rm and}\quad{\bf c\cdot
d}=(s_c t_d - t_c s_d)^2,\label{ecu52wd}
\end{equation}
whence \begin{equation} ({\bf c\cdot d})^{\lambda_b/2}({\bf b\cdot
c})^{\lambda_d/2}({\bf b\cdot  d})^{\lambda_c/2}= (s_c t_d - t_c
s_d)^{\lambda_b} (s_b t_c - t_b s_c)^{\lambda_d} (s_b t_d - t_b
s_d)^{\lambda_c}. \label{ecu52we}
\end{equation}
By binomially expanding $(s_b t_c - t_b s_c)^{\lambda_d}$,
\begin{equation}
(s_b t_c-t_b s_c)^{\lambda_d}=\sum_{j=0}^{\lambda_d}
\frac{(-1)^{\lambda_d-j}\lambda_d!}{(\lambda_d-j)!j!}s_b^{j}
t_b^{\lambda_d-j}t_c^{j}s_c^{\lambda_d-j},\label{ecu52wf}
\end{equation}
we see that the only term in the binomial expansion of $(s_b t_d -
t_b s_d)^{\lambda_c}$ which multiplied by $s_b^j
t_b^{\lambda_d-j}$ in (\ref{ecu52wf}) yields $s_b^{\mu_b}
t_b^{\nu_b}\equiv \rho_b$ is that containing $s_b^k
t_b^{\lambda_c-k}$ with $k=-j+\mu_b$ which, in addition, also
makes $\lambda_c-k+\lambda_d-j=\lambda_c+\lambda_d-\mu_b=\nu_b$.
Therefore,
\begin{equation}
\left[(s_b t_d - t_b s_d)^{\lambda_c}(s_b t_c - t_b
s_c)^{\lambda_d}\right]_{\rho_b\equiv s_b^{\mu_b} t_b^{\nu_b}}=
\sum_{j=0}^{\lambda_d}
\frac{(-1)^{l_b-m_b}\lambda_d!\lambda_c!t_c^{j}s_c^{\lambda-j}t_d^{\mu_b-j}
t_d^{-\lambda_d+\nu_b+j}}{(\lambda_d-j)!j!(\mu_b-j)!(\nu_b-\lambda_d+j)!}
.\label{ecu52wg}
\end{equation}
Also, the term containing $s_c^k t_c^{\lambda_b-k}t_d^k
s_d^{\lambda_b-k}$ with $k=j-\lambda_d+\mu_c$ is the only one in
the binomial expansion of $(s_c t_d - t_c s_d)^{\lambda_b}$ which
can yield $s_c^{\mu_c} t_c^{\nu_c}s_d^{\mu_d} t_d^{\nu_d}\equiv
\rho_c\rho_d$ when multiplied by
$t_c^{j}s_c^{\lambda-j}t_d^{\mu_b-j}s_d^{-\lambda_d+\nu_b+j}$ in
(\ref{ecu52wg}). It is clear that such a value of $k$ also makes
$\lambda_b-k+j=\lambda_b+\lambda_d-\mu_c=\nu_c$ and, since we must
also have that $\nu_d=k+\mu_b-j=\mu_b-\lambda_d+\mu_c$, the
$m$-indexes must satisfy the condition
\begin{equation}
M=m_b + m_c + m_d = 0. \label{ecu52wh}
\end{equation}
That relation (\ref{ecu52wh}) must be satisfied for
(\ref{ecu52wb}) to be non vanishing becomes obvious if we express
the integral on the LHS in spherical coordinates, since it assures
the independence of the azimuthal coordinate of the integrand.
Observe that (\ref{ecu52wh}) also implies that the exponent of
$s_d$ is
$\lambda_b-k-\lambda_d+\nu_b+j=\lambda_b+\nu_b-\mu_c=\mu_d$, as it
should be. Therefore,
\begin{equation}
\begin{split}
&\left[({\bf c \cdot d})^{\lambda_b/2}({\bf b\cdot
c})^{\lambda_d/2}({\bf b\cdot
d})^{\lambda_c/2}\right]_{\rho_b\rho_c\rho_d}= \\
& \sum_{j=0}
\frac{(-1)^{l_c+l_d+m_b+j}\lambda_b!\lambda_c!\lambda_d!}{j!(\lambda_d-j)!(\mu_b-j)!(\nu_b-\lambda_d+j)!(j+\mu_c-\lambda_d)!(\nu_c-j)!}
.\label{ecu52wi}
\end{split}
\end{equation}
where the sum in (\ref{ecu52wi}) is over all integers $j$ for
which the factorials all have nonnegative arguments.Inserting
(\ref{ecu52wi}) into the RHS of (\ref{122cba}) and the later into
(\ref{ecu52wb}) we finally obtain, after dividing by $R^{L+2}$,
\begin{equation}
\int d\Omega Y_{l_bm_b}Y_{l_cm_c}Y_{l_dm_d}=
\sqrt{\frac{(2l_b+1)(2l_c+1)(2l_d+1)}{4\pi}}
\begin{pmatrix} l_b \hfill & l_c \hfill & l_d \hfill \\
0 \hfill & 0 \hfill & 0 \hfill
\end{pmatrix}\begin{pmatrix}
l_b \hfill & l_c \hfill & l_d \hfill \\
m_b \hfill & m_c \hfill & m_d \hfill
\end{pmatrix}
\label{ecu52wjjjj}
\end{equation}
where the integral in the LHS vanishes unless
\begin{equation}
m_b+m_c+m_d=0,\quad l_b+l_c\ge l_d,\quad l_c+l_d\ge l_b\quad {\rm
and}\quad l_d+l_b\ge l_c ,\label{ecu52wja}
\end{equation}
and we introduced the standard Wigner 3j-symbols\cite{Eric}
defined as
\begin{equation}
\begin{pmatrix}
l_b \hfill & l_c \hfill & l_d \hfill \\
m_b \hfill & m_c \hfill & m_d \hfill
\end{pmatrix}\equiv
\sum_{j=0}
\frac{(-1)^{l_b+l_c-m_d}\sqrt{\Delta(l_b,l_c,l_d)}\sqrt{\mu_b!\mu_c!\mu_d!\nu_b!\nu_c!\nu_d!}}
{j!(\lambda_d-j)(\mu_b-j)!(\nu_c-j)!(\nu_b-\lambda_d+j)!(\mu_c-\lambda_d+j)!}
,\label{ecu52wk}
\end{equation}
where
\begin{equation}
\Delta(l_b,l_c,l_d)\equiv\frac{\lambda_b!\lambda_c!\lambda_d!}{(L+1)!}
, \label{ecu52wj}
\end {equation}
and
\begin{equation}
\begin{pmatrix}
l_b \hfill & l_c \hfill & l_d \hfill \\
0 \hfill & 0 \hfill & 0 \hfill
\end{pmatrix}=\frac{(-1)^{L/2}\sqrt{\Delta(l_b,l_c,l_d)}(L/2)!}
{(\lambda_b/2)!(\lambda_c/2)!(\lambda_d/2)!}\,.
\end{equation}

\section{Conclusions.}
We have presented the coordinate-free approach to spherical
harmonics based on the contributions by Maxwell, Thomson and Tait.
It has been shown that many results of the theory of spherical
harmonics needed in physical applications can straightforwardly
and systematically derived from the simple properties of
elementary solid harmonics $({\bf b\cdot r})^l\,$ ($\,{\bf b}$
null vector), Maxwell's harmonics and Maxwell's integral theorem.
In particular, we have provided simple and apparently new proofs
of known results such as Maxwell's integral theorem and Legendre's
addition theorem, the latter being obtained in terms of a general
basis of orthonormal surface harmonics. Also, it has been shown
how the derivation of recursions relations can be made simpler and
more systematic using Maxwell's solid $V_l$-harmonics than - as is
conventionally done - using surface harmonics and spherical
coordinates. Surface harmonics expansion have  been discussed and
their relation with the method of images illustrated and, also,
they have been used to efficiently compute the interaction energy
of non-overlapping charge distributions. We have reviewed in a
unified manner partial wave expansions and provided a seemingly
new proof of Hobson's integral theorem. A procedure based on
compact vector methods has been given to express the elements of
the rotation matrix in a form which is independent of the
parametrization of the rotation. Finally, we have shown how
Hobson's integral theorem can be used to find integrals involving
products of three (or more) spherical harmonics.

\appendix
\section{Derivation of the expression for $Y_{lm}$ in spherical coordinates.}
The $V_{l}$-harmonic $V_{lm}$ defined by (\ref{ecu31}) can be
easily expressed in spherical coordinates through the familiar
Legendre polynomials as follows. First, consider the well known
Legendre's expansion
\begin{equation}
\frac{1}{\sqrt{1-2s\mu+s^2}}=\sum_{j=0}^{\infty} s^j P_j(\mu),
\label{ecu32ap}
\end{equation}
where $P_j(\mu)$ is the Legendre polynomial of degree $j$,
\begin{equation}
P_j(\mu)=\frac{1}{2^j\,j!}\frac{d^j}{d\mu^j}\big(\mu^2-1\big)^j.
\label{ecu33ap1}
\end{equation}
Taking $m$ times the derivative with respect to $\mu$ in both
sides of (\ref{ecu32ap}) we get
\begin{equation}
\frac{1}{(1-2s\mu+s^2)^{m+1/2}}=\frac{1}{(2m+1)!!}
\sum_{j=m}^{\infty} s^{j-m} \frac{d^m
P_j}{d\mu^m}=\frac{1}{(2m+1)!!} \sum_{j=0}^{\infty} s^{j}
\frac{d^m P_{j+m}}{d\mu^m}, \label{ecu34ap}
\end{equation}
where we have taken into account that $d^m P_j/d\mu^m=0$ for $j <
m$.  Making $s=h/r$ and $\mu=z/r={\bf e}_z\cdot {\bf r}/r$, with
$r=(x^2+y^2+z^2)^{1/2}$, we can write (\ref{ecu34ap}) as
\begin{equation}
\frac{1}{[x^2+y^2+(z-h)^2]^{m+1/2}}=
\frac{r^{-2m}}{(2m+1)!!}\sum_{j=0}^{\infty} \frac{h^j}{r^{j+1}}
\frac{d^m P_{j+m}}{d\mu^m}, \label{ecu35ap}
\end{equation}
Now, it is readily seen that we can compute $\partial^{l-m}
(1/r^{2m+1})/\partial z^{l-m}$ as $(-1)^{l-m}$ times the
$(l-m)$-th derivative of (\ref{ecu35ap}) with respect to $h$
evaluated at $h=0$, or
\begin{equation}
\frac{\partial^{l-m}}{\partial z^{l-m}}
\frac{1}{r^{2m+1}}=(-1)^{l-m}\,\frac{(l-m)!}{(2m+1)!!}\,
r^{-(l+m+1)}\frac{d^m P_{l}}{d\mu^m}. \label{ecu36ap}
\end{equation}
Therefore,
\begin{equation}
V_{lm}({\bf r})=\frac{(-1)^{l}(l-m)!}{2^m r^{l+m+1}}\, \xi^m
\frac{d^m P_{l}}{d\mu^m}=\frac{(-1)^{l}(l-m)!}{2^m r^{l+1}}\,
\sin^m \theta e^{im\varphi} \frac{d^m P_{l}}{d\mu^m}.
\label{ecu37ap}
\end{equation}
Substituting (\ref{ecu37ap}) into (\ref{ecu45}) and using
(\ref{ecu33ap1}) we obtain the familiar\cite{But,Jack,Fe}
expression for $Y_{lm}$
\begin{equation}
Y_{lm}({\bf r})= \sqrt{\frac{2l+1}{4\pi}}
\sqrt{\frac{(l-m)!}{(l+m)!}} \frac{(-1)^m}{2^l l!} \sin^m \theta
e^{im\varphi} \frac{d^{l+m} (\mu^2-1)^l}{d\mu^{l+m}},
\label{ecu38ap}
\end{equation}
which is valid also for $m<0$.

\begin{section}{Acknowledgements}
The author is indebted to Professors J. M. Gordillo and  A.
Barrero for helpful discussions and continuous encouragement
during the writing of the paper.
\end{section}


\begin{thebibliography}{5}

\bibitem{But} G.B. Arfken and H.J. Weber, \textsl{Mathematical Methods for Physicists} (Academic Press, San Diego, 2001)

\bibitem{Jack} J.D. Jackson, \textsl{Classical Electrodynamics} (Wiley, New York, 1999)

\bibitem{TT} W.T. Thomson and P.G. Tait,\textsl{Treatise on Natural Philosophy. 2 Vols.} (Dover, New York, 1962)

\bibitem{Max} J.C. Maxwell, \textsl{A Treatise on Electricity and Magnetism} (Dover, New York, 1954)

\bibitem{Hob} E.W. Hobson, \textsl{The Theory of Spherical and Ellipsoidal Harmonics} (University Press, Cambridge, 1932)

\bibitem{Niven} W.D. Niven, Phil. Trans. Roy. Soc. {\bf 170}, 379-416  (1879)

\bibitem{Abra} M. Abramowitz and J.A. Stegun, \textsl{Handbook of Mathematical Functions} (Dover, New York, 1965)

\bibitem{Fe} N.M. Ferrers, \textsl{An Elementary Treatise on Spherical Harmonics} (Macmillan, London, 1877)

\bibitem{Wenig2} E.J. Weniger, Int. J. Quantum Chem. {\bf 90}, 92-104 (2002)

\bibitem{PerSab}  M. P\'{e}rez-Saborid , arxiv:math-ph/0806.3367  (2010)


\bibitem{Apple} J. Applequist, Theor. Chem. Acc. {\bf 107},
103-115 (2002)

\bibitem{Weeks}  J. R. Weeks, arxiv:astro-ph/0412231 v2  (2004)

\bibitem{Dowker} J. S. Dowker, arXiv:math-ph/0805.1904 v1  (2008)


\bibitem{Luca} L. Bombelli, \textsl{Theoretical Physics, the Universe, and
Everything} http://www.phys.olemiss.edu/$\sim$
luca/Topics/s/spher-harm.htlm, 2008)

\bibitem{Kra} H.A. Kramers, \textsl{Quantum Mechanics} (Dover, New York, 1964)

\bibitem{Rokhlin} V. Rokhlin, J. Comp. Phys. {\bf 86} 414-439
(1990)

\bibitem{Rusos} N. A. Gumerov and R. Duraiswami \textsl{Fast multipole methods for the Helmholtz equation in three dimensions.} (Elsevier, Amsterdam, 2004)


\bibitem{Sch} J. Schwinger, L. Deraad, K. Milton, W-Y. Tsai, \textsl{Classical Electrodynamics} (Westview Press, 1998)

\bibitem{Heit} W. Heitler , \textsl{The Quantum Theory of Radiation} (Dover, New York, 1984)

\bibitem{Ray} Lord Rayleigh, Proc.  London Math. Soc. {\bf 4}, 253 (1873)

\bibitem{Rowe} E.G. Peter Rowe, J. Math. Phys. {\bf 19}, 1962 (1978)

\bibitem{Bied} L.C. Biedenharn and J.D. Louck \textsl{Angular Momentum in Quantum Physics} (Adisson-Wesley, Reading, 1981)

\bibitem{Wig} E.P. Wigner, \textsl{Group Theory} (Academic, New York, 1959)

\bibitem{Mana1} N.L. Manakov, A.V. Meremianin, A.F. Starace, Phys.Rev. A {\bf 57} 3233-3244 (1998)

\bibitem{Mana2} N.L. Manakov, A.V. Meremianin, A.F. Starace, Phys.Rev. A {\bf 64} 032105 (2001)

\bibitem{Mana3} N.L. Manakov, A.V. Meremianin, A.F. Starace, J.Phys.B: At. Mol. Opt. {\bf 35} 77-91 (2002)

\bibitem{Gaunt} J.A. Gaunt, Phil. Trans. Roy. Soc. {\bf 228}, 151-196  (1929)

\bibitem{Weyl} H. Weyl, \textsl{The Theory of Groups and Quantum Mechanics} (Dover, New York, 1950)

\bibitem{Eric} Eric W. Weisstein \textsl{"Wigner 3j-Symbol." From {\rm MathWorld}-A Wolfram Web Resource} (http://mathworld.wolfram.com/Wigner3j-Symbol.html)













\end{thebibliography}
\end{document}